\documentclass[epj]{svjour}
\usepackage{latexsym}
\usepackage{graphics}
\usepackage[latin1]{inputenc}
\usepackage{rotating}
\usepackage{amsmath}
\begin{document}
\renewcommand{\textfraction}{0.0}
\renewcommand{\floatpagefraction}{1.0}
\title{Photoproduction of {\boldmath{$\pi^0$}}-pairs off protons and off neutrons}
\author{
M.~Dieterle\inst{1},
M.~Oberle\inst{1},
J.~Ahrens\inst{2},		
J.R.M.~Annand\inst{3},	   
H.J.~Arends\inst{2},
K.~Bantawa\inst{4},
P.A.~Bartolome\inst{2},	    
R.~Beck\inst{5},	 
V.~Bekrenev\inst{6},
H.~Bergh\"auser\inst{7},	       
A.~Braghieri\inst{8},  	 
D.~Branford\inst{9},	     
W.J.~Briscoe\inst{10},	      
J.~Brudvik\inst{11},		
S.~Cherepnya\inst{12},
S.~Costanza\inst{8}    	   
B.~Demissie\inst{10},
E.J.~Downie\inst{2,3,10},	    
P.~Drexler\inst{7}, 
L.V. Fil'kov\inst{12}, 
A.~Fix\inst{13},
S.~Garni\inst{1},
D.I.~Glazier\inst{3},	
D. Hamilton\inst{3},      
E.~Heid\inst{2},
D.~Hornidge\inst{14}, 
D.~Howdle\inst{3}, 
G.M.~Huber\inst{15},		
O.~Jahn\inst{2},
T.C.~Jude\inst{9},
A. K{\"a}ser\inst{1},  	      
V.L.~Kashevarov\inst{12,2},
I. Keshelashvili\inst{1}\thanks{present address: Institut f\"ur Kernphysik, Forschungszentrum J\"ulich,
52425 Jülich, Germany},  
R.~Kondratiev\inst{16},	  
M.~Korolija\inst{17},  
B.~Krusche\inst{1},
V.~Lisin\inst{16},		  
K.~Livingston\inst{3},	   
I.J.D.~MacGregor\inst{3},	   
Y.~Maghrbi\inst{1},
J.~Mancell\inst{3},  
D.M.~Manley\inst{4}, 
Z.~Marinides\inst{10},	      
J.C.~McGeorge\inst{3}, 
E.~McNicoll\inst{3}, 	 
D.~Mekterovic\inst{17},	  
V.~Metag\inst{7},
S.~Micanovic\inst{17},
D.G.~Middleton\inst{14},
A.~Mushkarenkov\inst{8},		
A.~Nikolaev\inst{5},   
R.~Novotny\inst{7},
M.~Ostrick\inst{2},
P.~Otte\inst{2},
B.~Oussena\inst{2,10},
P.~Pedroni\inst{8}, 
F.~Pheron\inst{1},	       
A.~Polonski\inst{16},  	  
S.~Prakhov\inst{2,10,11}, 
J.~Robinson\inst{3}, 		   
T.~Rostomyan\inst{1},	       
S.~Schumann\inst{2},
M.H.~Sikora\inst{9},	     
D.I.~Sober\inst{18}, 	      
A.~Starostin\inst{11},
Th.~Strub\inst{1},		
I.~Supek\inst{17},		  
M.~Thiel\inst{7},	    
A.~Thomas\inst{2},		 
M.~Unverzagt\inst{2,5},
N.K.~Walford\inst{1},			  
D.P.~Watts\inst{9},
D.~Werthm\"{u}ller\inst{1}\thanks{present address: SUPA School of Physics and Astronomy, University of Glasgow,
G12 8QQ, United Kingdom},
L.~Witthauer\inst{1}
\newline(The A2 Collaboration)
\mail{B. Krusche, Klingelbergstrasse 82, CH-4056 Basel, Switzerland,
\email{Bernd.Krusche@unibas.ch}}
}

\institute{Department of Physics, University of Basel, CH-4056 Basel, Switzerland
  \and Institut f\"ur Kernphysik, University of Mainz, D-55099 Mainz, Germany
  \and SUPA School of Physics and Astronomy, University of Glasgow, G12 8QQ, United Kingdom
  \and Kent State University, Kent, Ohio 44242, USA
  \and Helmholtz-Institut f\"ur Strahlen- und Kernphysik, University of Bonn, D-53115 Bonn, Germany
  \and Petersburg Nuclear Physics Institute, RU-188300 Gatchina, Russia
  \and II. Physikalisches Institut, University of Giessen, D-35392 Giessen, Germany
  \and INFN Sezione di Pavia, I-27100 Pavia, Pavia, Italy
  \and SUPA School of Physics, University of Edinburgh, Edinburgh EH9 3JZ, United Kingdom
  \and Center for Nuclear Studies, The George Washington University, Washington, DC 20052, USA
  \and University of California Los Angeles, Los Angeles, California 90095-1547, USA
  \and Lebedev Physical Institute, RU-119991 Moscow, Russia
  \and Laboratory of Mathematical Physics, Tomsk Polytechnic University, Tomsk, Russia
  \and Mount Allison University, Sackville, New Brunswick E4L1E6, Canada
  \and University of Regina, Regina, SK S4S-0A2 Canada
  \and Institute for Nuclear Research, RU-125047 Moscow, Russia
  \and Rudjer Boskovic Institute, HR-10000 Zagreb, Croatia
  \and The Catholic University of America, Washington, DC 20064, USA 
}
\authorrunning{M. Dieterle et al.}
\titlerunning{Photoproduction of $\pi^0$ pairs off protons and neutrons}

\abstract
{Total cross sections, angular distributions, and invariant-mass distributions have been 
measured for the photoproduction of $\pi^0\pi^0$ pairs off free protons and off nucleons
bound in the deuteron. The experiments were performed at the MAMI accelerator facility 
in Mainz using the Glasgow photon tagging spectrometer and the Crystal Ball/TAPS detector. 
The accelerator delivered electron beams of 1508 and 1557~MeV, which produced 
bremsstrahlung in thin radiator foils. The tagged photon beam covered energies up to 1400~MeV. 
The data from the free proton target are in good agreement with previous measurements and were 
only used to test the analysis procedures. The results for differential cross sections 
(angular distributions and invariant-mass distributions) for free and quasi-free protons are 
almost identical in shape, but differ in absolute magnitude up to 15\%. Thus, moderate final-state 
interaction effects are present. The data for quasi-free neutrons are similar to the proton 
data in the second resonance region (final state invariant masses up to  $\approx$1550~MeV), 
where both reactions are dominated by the $N(1520)3/2^-\rightarrow \Delta(1232)3/2^+\pi$ decay. At higher 
energies, angular and invariant-mass distributions are different. A simple analysis of the shapes 
of the invariant-mass distributions in the third resonance region is consistent with strong 
contributions of an $N^{\star}\rightarrow N\sigma$ decay for the proton, while the reaction 
is dominated by a sequential decay via a $\Delta\pi$ intermediate state for the neutron. 
The data are compared to predictions from the Two-Pion-MAID model and the Bonn-Gatchina 
coupled channel analysis. 
} 
\maketitle

\section{Introduction}
The properties of the nucleon and its excited states are a key for the investigation of the 
strong interaction in the non-perturbative regime. There are several new developments on the theory 
side. Fully relativistic quark-model approaches have been developed \cite{Plessas_13} and also 
the direct application of the fundamental properties of Quantum Chromodynamics (QCD) to nucleon 
structure has made significant progress. The application of the Dyson-Schwinger approach 
to QCD has led to promising results (see e.g. \cite{Chen_12,Eichmann_12,Aznauryan_13}) and 
the advances in lattice gauge calculations allowed first predictions of the excitation 
spectrum based on unquenched lattice simulations \cite{Edwards_11}. These results are still in 
early stages, using pion masses around 400 MeV (lattice predictions for ground-state properties
are nowadays possible for physical quark masses). However, they are interesting because 
they `re-discovered' the SU(6)$\otimes$O(3) excitation structure of the nucleon with 
a level counting consistent with the standard non-relativistic quark model.

These successes are complemented by experimental efforts using photon-induced meson production
reactions for the study of nucleon resonances. They aim at a more complete and reliable database 
for the nucleon excitation spectrum. The comparison of predictions and experimental data is 
unsatisfactory, in particular for center-of-mass energies above $\approx$ 1800~MeV, where a large
discrepancy between predicted and observed level density exists \cite{PDG}, which is known
as the `missing resonance' problem. With a few exceptions, for most quantum numbers only the lowest 
lying state has been identified experimentally \cite{PDG}, while quark models predict many more 
states at higher excitation energies. A possible reason could be experimental bias.
A few years ago the {\it Review of Particle Physics} (RPP) listed only nucleon resonances that had 
been identified in elastic and inelastic pion scattering reactions. Only in the two most recent 
updates \cite{PDG,PDG_12} were states `established' by observations in photon-induced reactions 
included. Possible bias from elastic pion scattering is obvious, states that do not significantly 
couple to $N\pi$ are suppressed in the initial and final state of this reaction. Such bias probably 
grows with excitation energy. An obvious reason is the increasing phase space for the emission of 
heavier mesons or meson pairs. However, the internal structure of the states may also play a role. 
One can, for example, expect that resonances that have more than one oscillator excited may tend 
to de-excite step-by-step via cascades involving intermediate states with only one oscillator excited
\cite{Thiel_15}. In this case, entire multiplets of resonances may contribute only weakly to 
single meson production.    

The recent experimental efforts tried to remove such bias by a large-scale study of photon-induced
meson production reactions. These experiments include measurements of sequential resonance decays
via $R\rightarrow R'\pi\rightarrow N\pi\pi$ ($R$, $R'$ nucleon resonances, $N=n,p$) decay chains. 
The corresponding final states are multiple meson production reactions. So far, $\pi\pi$ 
(see e.g. \cite{Thiel_15,Braghieri_95,Krusche_03,Sarantsev_08,Thoma_08,Kashevarov_12,Zehr_12} 
and Refs. therein) and $\pi\eta$ 
(see e.g. \cite{Ajaka_08,Horn_08b,Kashevarov_09,Kashevarov_10,Gutz_14,Kaeser_15} 
and Refs. therein) pairs have been studied. The analysis of such reactions is challenging and requires 
the measurement of several observables. The reaction amplitudes for photoproduction 
of single pseudoscalar mesons can be completely fixed by the measurement of at least eight carefully 
chosen observables \cite{Chiang_97} as functions of two independent kinematic variables. 
However, for pseudoscalar meson pairs \cite{Roberts_05}, the measurement of eight observables as 
functions of five kinematic parameters fixes only the magnitude of the amplitudes and 15 observables 
would be necessary to extract the complex phases also. `Complete experiments' are therefore not 
practical. Nevertheless, current efforts aim at measurements of invariant-mass distributions 
(of meson-meson and meson-nucleon pairs), angular distributions and at least some canonical single 
(such as $\Sigma$, $T$, $P$, $I^{\odot}$, $I^{C}$, $I^S$...) and double (such as $G$, $F$, $H$...) 
polarization observables.   

\subsection{Photoproduction of {\boldmath{$\pi^0\pi^0$}} pairs}
Among the different final states, $\pi^0$ pairs play a special role. Although their production
cross section throughout the second and third nucleon resonance regions is not as large as for
mixed-charge or double-charge pairs, it is still sizeable (on the order of 5 - 10 $\mu$b). Their  
advantage over the other isospin channels is the suppression of non-re\-so\-nant background terms.
Contributions from the direct coupling of the incident photon to the charge of the mesons, e.g. 
in the $t$-channel, are large for final states with charged pions, but do not contribute to the
production of $\pi^0$ pairs. In addition, production of $\rho$-mesons cannot contribute
because the $\rho^0$ decays into $\pi^+\pi^-$, but not into $\pi^0\pi^0$. Therefore, this 
final state is ideally suited for the investigation of sequential decays of $s$-channel 
nucleon resonance excitations.   

The total cross section (see e.g. \cite{Sarantsev_08,Kashevarov_12}) shows a pronounced double-hump 
structure with two bumps corresponding to the second and third nucleon resonance regions and the 
invariant-mass spectra of the pion-nucleon pairs indicate significant contributions
from the $\pi^0 \Delta(1232)$ intermediate state. Different observables for the photoproduction 
of $\pi^0$ pairs off the proton were investigated 
in the past for the second resonance region composed of the $N(1440)1/2^+$, $N(1535)1/2^-$, 
and $N(1520)3/2^-$ nucleon states with the DAPHNE \cite{Braghieri_95,Ahrens_05}, 
TAPS \cite{Haerter_97,Wolf_00,Sarantsev_08}, and Crystal Ball/TAPS 
\cite{Kashevarov_12,Zehr_12,Kotulla_04,Krambrich_09,Oberle_13}
detectors at the MAMI accelerator in Mainz. Two of those experiments \cite{Kashevarov_12,Oberle_13}
also covered the energy range of the third nucleon resonance region for which further measurements
have been reported from the GRAAL experiment \cite{Assafiri_03} and from the Crystal Barrel/TAPS
setup at ELSA \cite{Thiel_15,Sarantsev_08,Thoma_08,Sokhoyan_15a,Sokhoyan_15b}. 

This reaction has been analyzed in the framework of different models. Surprisingly, even for low 
incident photon energies in the second resonance region, where only a few resonances contribute,
the results diverged. Analyses in the framework of the `Valencia' model
\cite{Gomez_96,Nacher_01,Nacher_02}, and also the Two-Pion-MAID model \cite{Fix_05}
emphasized a strong contribution of the 
$N(1520)3/2^-\rightarrow \pi^0\Delta(1232)3/2^+\rightarrow \pi^0\pi^0 p$ reaction chain. 
The work by Murphy and Laget discussed in the paper with the GRAAL results \cite{Assafiri_03} 
proposed a dominant contribution of the $N(1440)1/2^+\rightarrow N\sigma$ decay.
However, this solution was later excluded by the experimentally established dominance of the 
$\sigma_{3/2}$ component in the total cross section measured with circularly polarized photons 
incident on longitudinally polarized protons \cite{Ahrens_05}. All above-mentioned
models have only small or even negligible contributions from the $\Delta (1700)3/2^-$ state, 
but the Bonn-Gatchina (BnGa) coupled channel analysis \cite{Sarantsev_08,Thoma_08}
claimed a substantial contribution of this state resulting in the double-hump
structure of the total cross section due to the interference of the excitation
of this resonance and the $N(1520)3/2^-$ state. A detailed analysis of the angular
distributions  reported by Kashevarov {\it et al.} \cite{Kashevarov_12} revealed significant 
contributions from $J=3/2$ partial waves for excitation energies below the $N(1520)3/2^-$ 
resonance. The nature of these contributions is still disputed. The $\Delta(1700)3/2^-$ would be broad 
enough to contribute already at these energies (as suggested by the BnGa analysis), 
but also $\pi^+\pi^-\rightarrow \pi^0\pi^0$ rescattering effects discussed in \cite{Kashevarov_12}, 
which are neglected in most models, could play a role. However, it is obvious that some 
significant contribution in this energy range must be missing in the Two-Pion-MAID model
\cite{Fix_05} since the agreement with the measured total cross section is poor in the threshold
region \cite{Zehr_12}. 

More ambiguities exist at higher incident photon energies, but recently rapid progress was made.
In the framework of the BnGa analysis, resonance contributions in the third nucleon resonance region 
were discussed in \cite{Thoma_08} and in the fourth resonance region and beyond in 
\cite{Thiel_15,Sokhoyan_15a,Sokhoyan_15b}. For the $\gamma p\rightarrow \pi^0\pi^0 p$ reaction, 
new data for cross sections and polarization observables measured with a linearly 
polarized photon beam were recently reported from the ELSA experiment and were used for a 
detailed re-analysis of this reaction with the BnGa model \cite{Sokhoyan_15b}.   

The investigation of the isospin degree of freedom, i.e. the disentanglement of contributions 
from $\Delta$ and $N^{\star}$ resonances, requires data for the $\gamma n\rightarrow \pi^0\pi^0 n$ 
reaction, which is only accessible with neutrons bound in light nuclei, in particular in the deuteron.
The only cross-section data available so far for the deuteron are two inclusive measurements of the 
$\gamma d\rightarrow \pi^0\pi^0 np$ reaction up to the second resonance region with TAPS at 
MAMI \cite{Krusche_99,Kleber_00} and an exclusive measurement of the quasi-free 
$\gamma n\rightarrow \pi^0\pi^0 n$ reaction throughout the second and third resonance region
by the GRAAL experiment \cite{Ajaka_07}. The latter \cite{Ajaka_07} also measured the beam 
asymmetry $\Sigma$ with a linearly polarized photon beam. Furthermore, results for the 
beam-helicity asymmetry $I^{\odot}$ for the quasi-free reaction off neutrons measured with a 
circularly polarized photon beam were recently reported in \cite{Oberle_13}. The results from the
GRAAL experiment \cite{Ajaka_07} for the total cross section were similar to the 
$\gamma p\rightarrow \pi^0\pi^0 p$ reaction in the second resonance region, but showed a significant
enhancement of the third resonance bump for the neutron. Different behavior in this energy range 
was expected because, due to the relevant photon couplings, the $N(1680)5/2^+$ state and the
$N(1675)5/2^-$ should make strong and much different contributions (the $N5/2^+$ dominating
for protons and the $N5/2^-$ for neutrons). Here, it was more surprising that the beam-helicity
asymmetries \cite{Oberle_13} are almost identical for the proton and neutron target, contradicting
the only available model prediction from the Two-Pion-Maid model.

In this paper, we report the results of a detailed study of the total cross section, the invariant-mass 
distributions, and the angular distributions of quasi-free photoproduction of $\pi^0$ pairs
from nucleons bound in the deuteron compared to the same observables for this final state measured
off free protons. The aim of this measurement was to study the isospin structure of resonances in the 
second and third resonance region and to further explore the techniques of the extraction of 
`almost free' neutron data also from this final state.
For the reactions of quasi-free nucleons, effects from nuclear Fermi motion
were eliminated (apart from unavoidable resolution effects, see e.g. \cite{Werthmueller_13,Werthmueller_14}) 
by a complete kinematic reconstruction of the final state as discussed in \cite{Krusche_11}.
The comparison of free and quasi-free proton data serves as a cross check for the influence of
nuclear effects or final-state interactions (FSI), which might obscure the properties of the 
elementary reaction off free nucleons for quasi-free reactions off bound nucleons. The importance
of such effects is not yet well under control in reaction models and can be quite different 
depending on the investigated final state. Recently studied examples are photoproduction of
$\pi^0$ and $\eta$ mesons. In the first case, substantial FSI effects were reported \cite{Dieterle_14},
while for $\eta$ production \cite{Werthmueller_14}, FSI effects were negligible (almost no deviations 
between free and quasi-free proton results above the level of systematic uncertainty). 
For the $\gamma N\rightarrow \pi^0\pi^0 N$ reaction no significant FSI effects were observed
for the beam-helicity asymmetries, as reported by Oberle {\it et al.} \cite{Oberle_13} (these results
were based on the same data set analyzed in this work). This fact does not exclude that absolute 
cross sections might be more strongly affected. 

\section{Experimental setup}
The data were measured at the tagged photon beam \cite{Anthony_91,Hall_96,McGeorge_08} 
of the Mainz MAMI accelerator \cite{Herminghaus_83,Kaiser_08} with a detector setup
combining the Crystal Ball \cite{Starostin_01} and TAPS \cite{Novotny_91,Gabler_94}  
electromagnetic calorimeters. Data from three beam times using a liquid deuterium target were used 
(beam times used different triggers, different target lengths, and different electron beam energies) 
and additionally one beam time with a liquid hydrogen target was analyzed; the parameters of these 
beam times are summarized in Table \ref{tab:beam}. 
\begin{table}[thh]
\begin{center}
  \caption[Summary of data sets]{
    \label{tab:beam}
     Main parameters of the data samples. 1st column: Target type (LD$_2$: liquid deuterium, 
     $\rho_d$ = 0.169 g/cm$^3$; LH$_2$: liquid hydrogen, $\rho_H$ = 0.071 g/cm$^3$), 
     2nd column: target length [cm], 3rd column: target surface density $\rho_s$ [nuclei/barn], 
     4th column: electron beam energy $E_{e^-}$ [MeV], 5th column:
     trigger conditions (multiplicity $M$ and energy sum in CB $E_{\Sigma}$[MeV]).
}
\vspace*{0.3cm}
\begin{tabular}{|c|c|c|c|c|}
\hline
Target & L[cm] & $\rho_s$[barn$^{-1}]$ & 
$E_{e^-}$[MeV] & $M$, $E_{\Sigma}$[MeV]\\
\hline\hline
  LD$_2$ & 4.72 & 0.231$\pm$0.005 & 1508 & $M2+$, $300$\\
  LD$_2$ & 4.72 & 0.231$\pm$0.005 & 1508 & $M3+$, $300$\\
  LD$_2$ & 3.02 & 0.147$\pm$0.003 & 1557 & $M2+$, $300$\\
  LH$_2$ & 10   & 0.422$\pm$0.008 & 1557 & $M3+$, $300$\\  
\hline
\end{tabular}
\end{center}
\end{table}
The previously published results for total cross 
sections and angular distributions for the $\gamma N\rightarrow N\eta$ \cite{Werthmueller_14} and the 
$\gamma N\rightarrow \pi^0 N$ reactions \cite{Dieterle_14}, the beam-helicity asymmetries
for the $\gamma N\rightarrow \pi^0\pi^0 N$ \cite{Oberle_13} and $\gamma N\rightarrow \pi^0\pi^{\pm} N$ 
\cite{Oberle_14} reactions, and the data for production of $\eta\pi$ pairs \cite{Kaeser_15} are 
based on the same measurements with the deuterium targets and many experimental details are given 
in the corresponding publications. Therefore, we summarize them only briefly here.   

The measurements were done with electron beam currents up to 20~nA and with electron beam energies
of 1508 or 1557~MeV. The bremsstrahlung photons were produced in a 10$\mu$m copper radiator
and the scattered electrons were momentum analyzed with the Glasgow photon tagger 
\cite{Anthony_91,Hall_96,McGeorge_08} covering photon energies up to 94\% of the incident 
electron energies with a typical resolution of 4~MeV (defined by the width of the focal plane
detectors; intrinsic resolution of the dipole magnet is much better). The electron beam was
longitudinally polarized so that the photon beam was circularly polarized. This degree of freedom
was used to extract beam-helicity asymmetries of three-body final states \cite{Oberle_13,Oberle_14},
but is not relevant for the present results. For all measurements, the tagger focal-plane counters 
for photon energies in the $\Delta$ resonance region (below $\approx$400~MeV) were deactivated. 
The experiments aimed at meson production in the second and third nucleon resonance region and 
the high count rates in the focal plane detectors corresponding to smaller incident photon energies 
would have limited the usable beam currents. The size of the photon beam was defined by a 4~mm 
collimator, which restricted the beam spot on the production target to a diameter of $\approx$1.3~cm.
The targets were Kapton cylinders of 4.76~cm length (3.02~cm for one of the deuterium runs) and
a diameter of $\approx$4~cm filled with liquid deuterium or liquid hydrogen (target densities
are given in Table \ref{tab:beam}). Background contributions from the target windows
($2 \times 125\ \mu$m Kapton) were determined with empty target runs and subtracted. 

\begin{figure}[thb]
\centerline{\resizebox{0.48\textwidth}{!}{%
  \includegraphics{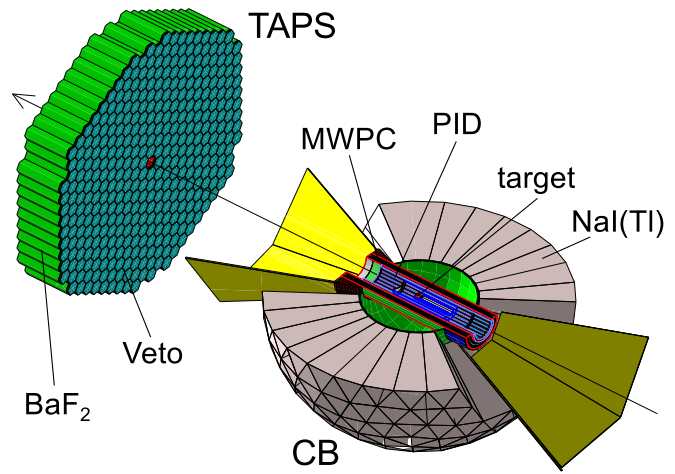}
}}
\caption{Experimental setup of Crystal Ball (only bottom hemisphere shown)
with PID detector and TAPS forward wall.
}
\label{fig:setup}       
\end{figure}

The mesons and the recoil nucleons were detected with a composite electromagnetic calorimeter
that covered almost the full solid angle. The target was mounted in the center of the Crystal Ball
(CB) detector \cite{Starostin_01}, which is schematically shown in Fig.~\ref{fig:setup}. This
detector is composed of 672 NaI(Tl) crystals with $15.7$ radiation lengths. It covers almost the full
azimuthal angle for polar angles between 20$^{\circ}$ and 160$^{\circ}$ (the detector has two 
hemispheres such that a small range of azimuthal angles in the horizontal plane is inactive material).
The forward angular range (polar angles between $5^{\circ}$ and $21^{\circ}$) was covered by the TAPS
detector \cite{Novotny_91,Gabler_94} with 384 BaF$_2$ crystals of 12 radiation lengths placed 
1.46~m (front face) downstream of the target. Charged particle identification was achieved with
a Particle Identification Detector (PID) \cite{Watts_04} mounted around the target inside the CB
and the TAPS Charged Particle Veto (CPV). The PID consists of 24 strips of plastic scintillator 
(50~cm long, 4~mm thick) arranged around the beam pipe. The CPV modules are 5~mm thick plastic scintillators
of hexagonal shape with identical geometry as the TAPS crystal front faces. In addition to simple 
hit/no hit patterns, both charged particle detector systems can be used together with the respective 
calorimeters for particle identification (e.g. proton-charged-pion separation) using the $\Delta E-E$ 
technique (for more details, see \cite{Werthmueller_14}). However, due to low light outputs 
from the CPV, its energy resolution was not very good. Therefore, in the present analysis
the $\Delta E-E$ particle identification was only used for the PID.     

The modules of the CB and the TAPS detector are equipped with two different discriminator systems. 
For the CB two leading-edge (LED) discriminators per crystal and for TAPS one LED and one 
constant-fraction (CFD) discriminator are used. One discriminator system (in case of TAPS the LED) 
serves for trigger generation. The event triggers, which were not identical for all beam times, 
were based on two conditions. The first condition defined the `hit' multiplicity in the detector, 
which approximated the number of particles (including photons) in an event. For this trigger 
component, CB and TAPS were subdivided into logical sectors. The 672 crystals of the CB were grouped 
into 45 units each containing up to 16 neighboring crystals and TAPS was divided into six triangular 
sectors. If the signal from at least one crystal in a sector exceeded a threshold ($\approx$30~MeV in CB, 
$\approx$35~MeV in TAPS) that sector contributed to the event multiplicity. Minimum hit multiplicities 
of two or three (see Table~\ref{tab:beam}) were required for the different beam times. Since all 
events of interest for the present analysis had at minimum four candidates for photons, only events for 
which the `hit'-multiplicity condition was satisfied by the photons alone were accepted in the data 
analysis. This avoided systematic effects from the detection and energy deposition of recoil nucleons. 
The second condition was a threshold for the analog energy sum of all signals from the CB, set to 
300~MeV for all beam times in order to suppress the abundant events from single $\pi^0$
production in the $\Delta$ resonance peak. These trigger conditions were reflected 
in the Monte Carlo (MC) simulations of the detection efficiency. For events which satisfied the trigger
conditions the second discriminator system with much lower thresholds (2~MeV for CB and 3-4~MeV for TAPS)
generated the pattern of activated crystals from which energy and timing information was processed and
stored. 

\section{Data analysis}
The following reactions were analyzed: $\gamma p\rightarrow \pi^0\pi^0 p$ (photoproduction off free protons),
$\gamma d \rightarrow \pi^0\pi^0 p(n)$ (photoproduction off quasi-free protons bound in the deuteron),
and $\gamma d\rightarrow \pi^0\pi^0 n(p)$ (photoproduction off quasi-free neutrons bound in the 
deuteron), where the nucleon in parenthesis was an undetected spectator. The reactions off the
deuterium target were analyzed in coincidence with the participant nucleons. Therefore, events
with exactly four neutral and one charged hit (for $\pi^0\pi^0 p$) and events with exactly five
neutral hits (for $\pi^0\pi^0 n$) were selected. Events with additional hits were discarded 
as background. The data from the liquid hydrogen target were analyzed in two different ways. For the 
results labeled `inc' (inclusive), only detection of the two $\pi^0$ mesons was required. This is the 
same type of analysis as used for most previous measurements of the $\gamma p\rightarrow \pi^0\pi^0 p$ 
reaction. In addition, an analysis with coincident detection of the recoil proton (four neutral, 
one charged hit) was done. The results were obtained in the same way as for quasi-free production off 
protons bound in the deuteron. Comparison of these two analyses allows the estimation of systematic 
uncertainties related to the recoil nucleon detection.

\begin{figure}[htb]
\centering{
\centerline{\resizebox{0.50\textwidth}{!}{%
  \includegraphics{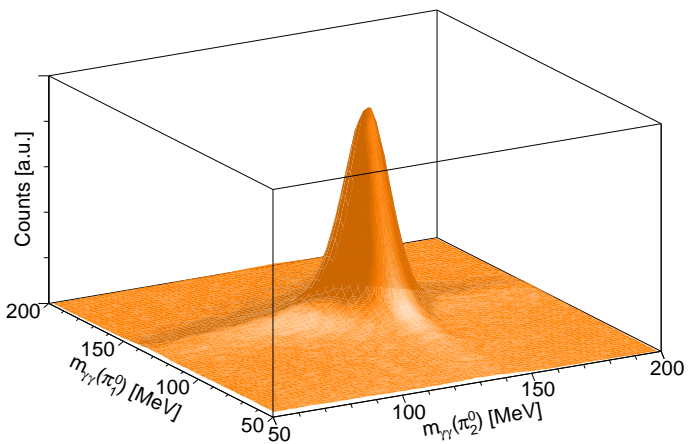}}
}}
\caption{Two-dimensional invariant-mass distribution of the `best' combination of four 
photons to two pairs for the $\gamma N\rightarrow \pi^0\pi^0 N$ reaction.
}
\label{fig:im2D}       
\end{figure}

\begin{figure}[thb]
\centerline{\resizebox{0.5\textwidth}{!}{%
  \includegraphics{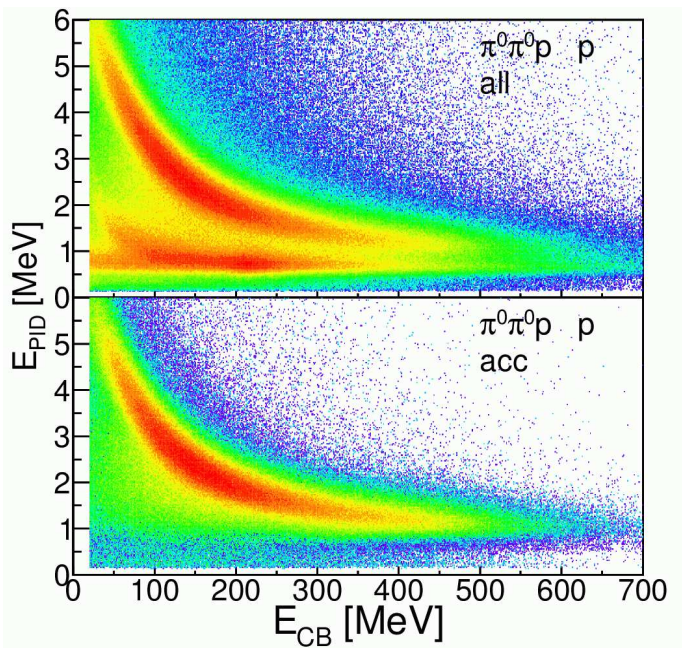}
}}
\caption{$\Delta E-E$ spectra from the CB-PID combination. Top: charged hits for all candidates
of the $\pi^0\pi^0 p$ final state (significant background structures from charged pions and electrons). 
Bottom: accepted candidates after kinematic cuts (no background structures).
}
\label{fig:ede}       
\end{figure}

The main steps in the analysis were (1) the identification of two neutral pions from their two-photon 
decays (achieved with invariant-mass analyses), (2) the identification of the recoil nucleons (hit 
pattern in PID, CPV, $\Delta E-E$ analysis, pulse-shape analysis (PSA), and time-of-flight (ToF) versus 
energy analysis), (3) removal of background from other reactions with two $\pi^0$ in the final state for 
example $\eta\rightarrow 3\pi^0$ (co-planarity and missing mass analyses), (4) kinematic reconstruction 
of the $W=\sqrt{s}$ of the $\pi^0\pi^0 N_p$ final state ($N_p$ = participant nucleon), and (5) 
absolute normalization of the cross sections from target density, photon flux, and instrumental 
detection efficiency. Since data from the same beam times were used to analyze 
different reactions or other observables for the same final states 
\cite{Kaeser_15,Oberle_13,Werthmueller_13,Werthmueller_14,Dieterle_14,Oberle_14}, all these 
steps and related systematic uncertainties were previously studied. Only the Monte Carlo simulations 
for the detection efficiency of the $\pi^0\pi^0 N_p$ final states had to be investigated in more detail 
than in \cite{Oberle_13} for the asymmetries (for asymmetries they cancel to a large extent so that
a more simple modeling could be used).

\begin{figure}[htb]
\centerline{\resizebox{0.5\textwidth}{!}{%
  \includegraphics{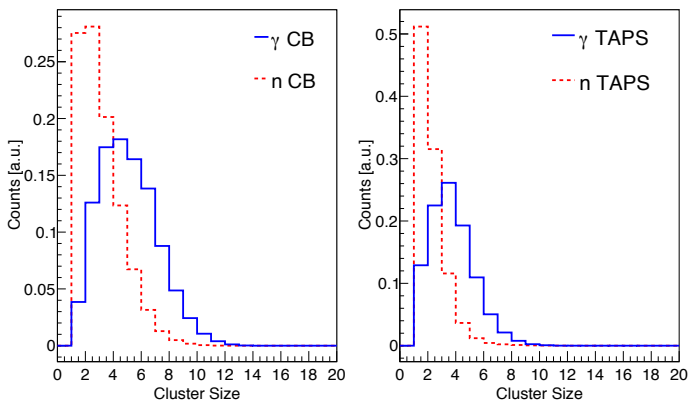}
}}
\caption{Cluster-size distributions for CB (left hand side) and TAPS (right hand side) for accepted photon
(solid blue histograms) and neutron (dashed red histograms) hits.
}
\label{fig:csd}       
\end{figure}

In the first analysis step, `hits' in the detector (i.e. connected clusters of energy depositions in the
scintillators) were classified as `neutral' or `charged' (see \cite{Werthmueller_14} for details)
depending on whether related hits in the PID (for CB) or in the CPV (for TAPS) were recorded. 
Subsequently, the invariant mass of photon pairs was analyzed. A minimum $\chi^2$ search was 
applied to all possible disjunct combinations of neutral hits to two pairs. The $\chi^2$ was defined 
by
\begin{equation}
\chi^2 = \sum_{i=1}^2 \left(\frac{m_{\gamma\gamma, i}-m_{\pi^0}}{\Delta m_{\gamma\gamma, i}}\right)^2 \;,
\end{equation}
where $m_{\gamma\gamma, i}$ are the invariant masses of the possible combinations of neutral 
hits to pion-decay photons, $\Delta m_{\gamma\gamma, i}$ are their uncertainties (calculated 
event-by-event from the known detector resolution), and $m_{\pi^0}$ is the nominal pion mass. 
A two-dimensional spectrum of the invariant masses of the `best' combinations (minimum $\chi^2$) 
is shown in Fig.~\ref{fig:im2D}. In the case of events with five neutral hits, the remaining cluster 
was treated as a neutron candidate. For neutral hits in the CB, this was the only criterion to 
distinguish photons from neutrons. Photon and neutron hits differ in average also in the cluster-size 
multiplicity (photons deposit energy in more crystals than neutrons). However, the cluster-size 
distributions overlap too much for an event-by-event separation of photons and neutrons. 
This has been studied for the same data sample for photoproduction of $\eta$-mesons \cite{Werthmueller_14}. 
Since $\eta$-mesons decay via $\eta\rightarrow\gamma\gamma$ and via $\eta\rightarrow 3\pi^0\rightarrow 6\gamma$ 
one has two data samples with different average photon energies and thus different cluster-size distributions.
However, the associated neutron candidates had identical cluster-size distributions, which is an 
additional test that no significant leakage of photons into the neutron sample occurs. Photons from the 
present $2\pi^0$ analysis have intermediate cluster sizes and the associated neutrons have again the 
same distribution as for the two $\eta$ decays. The overlapping cluster-size distributions of photon and 
neutron hits for the present analysis are shown in Fig.~\ref{fig:csd}. For the TAPS forward wall the excellent
separation of photon and neutron hits can be demonstrated with the PSA and ToF-versus-energy analysis. 

\begin{figure*}[thb]
\centerline{\resizebox{0.9\textwidth}{!}{%
  \includegraphics{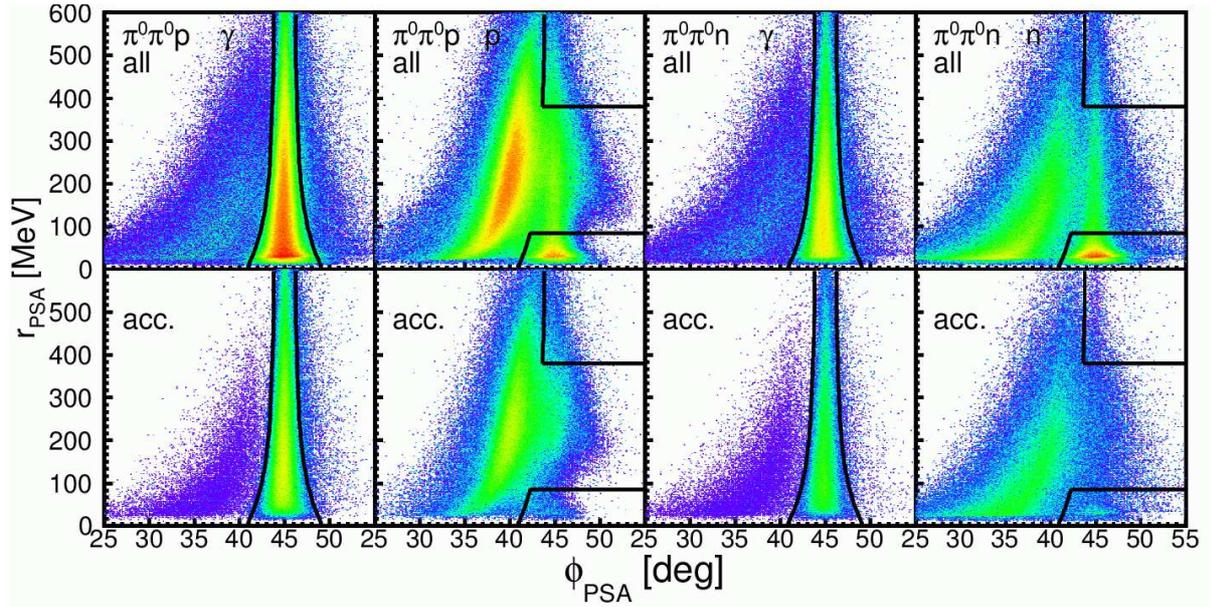}
}}
\caption{Results of TAPS PSA analysis. Top row: all events (assignment of neutral
hits to photon or neutron candidates by $\chi^2$ search of invariant masses), bottom row: events 
accepted after kinematic cuts. Black lines indicate PSA cuts. Accepted hits for photons between lines, 
accepted hits for recoil nucleons everything outside rectangles in corners of the spectra. 
From left to right: photon candidates for $\pi^0\pi^0 p$ final state, proton candidates 
for the same final state, photon candidates for $\pi^0\pi^0 n$ final states, and neutron candidates for 
the same final state.
}
\label{fig:tpsa}       
\end{figure*}

\begin{figure*}[hhtb]
\centerline{\resizebox{0.9\textwidth}{!}{%
  \includegraphics{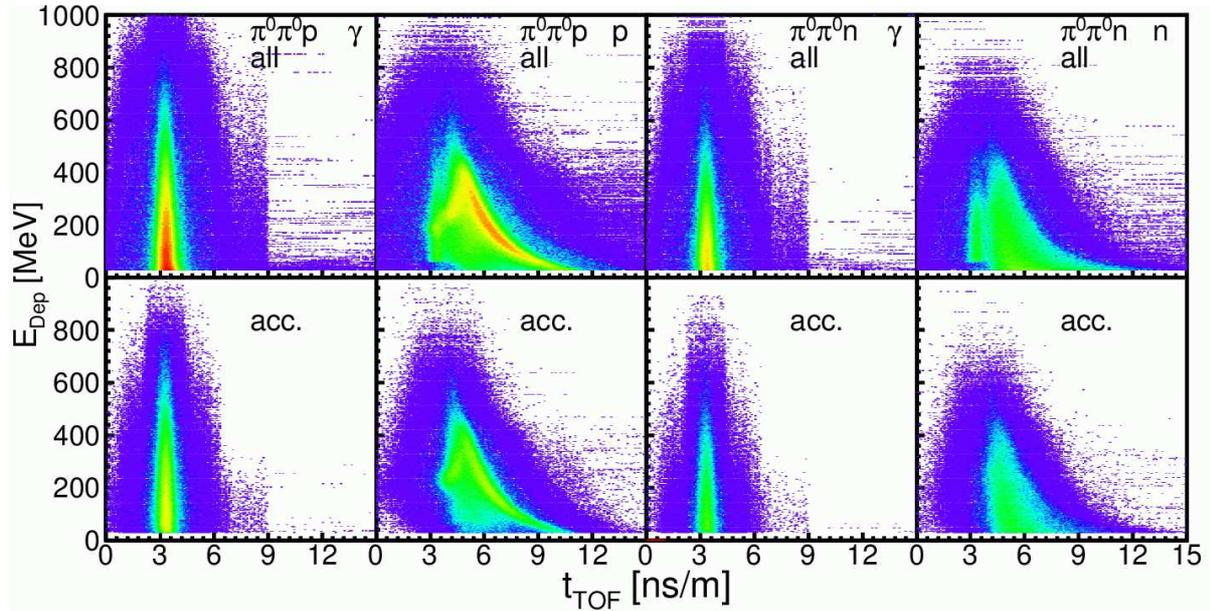}  
}}
\caption{ToF-versus-energy analysis for hits in TAPS. Rows and columns same as in Fig.~\ref{fig:tpsa}.
}
\label{fig:ttof}       
\end{figure*}

The results for further particle-identification methods like 
$\Delta E-E$ with CB and PID (separation of protons from charged pions) and PSA and ToF-versus-energy 
for TAPS are summarized in Figs.~\ref{fig:ede},\ref{fig:tpsa}, and \ref{fig:ttof}. These figures 
show the spectra for all events included in the invariant-mass analysis and the spectra for those 
events that also passed the subsequent selection steps. At this stage, cuts were only applied to the PSA 
information from TAPS (see \cite{Zehr_12} or \cite{Werthmueller_14} for details of PSA).
 
For further analysis, only events with both invariant masses within $\pm 3\sigma$ of the peak 
position (angle and energy dependent) were accepted. 
The small background structure below the peak was subtracted by a side-band analysis. This had to 
be done identically for the results of the Monte Carlo simulations (which are discussed at the end 
of this section) because part of the background is of combinatorial nature from true double 
$\pi^0$ events and thus contributes also to the simulations.


\begin{figure*}[thb]
\centerline{\resizebox{1.0\textwidth}{!}{%
  \includegraphics{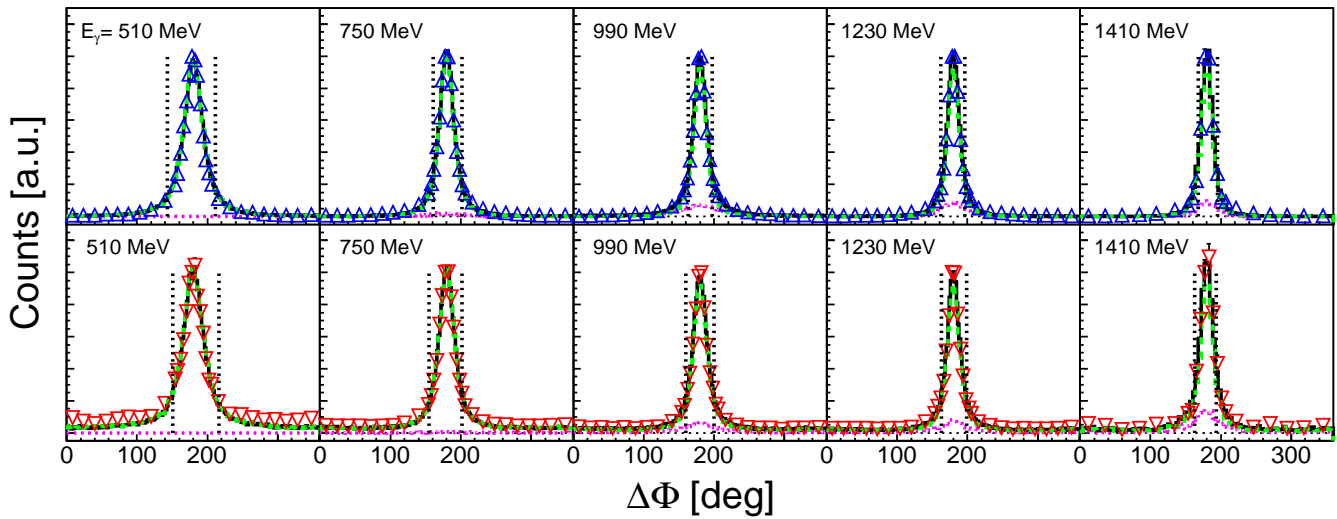}
}}
\caption{Spectra of the azimuthal difference between the three-momenta of the two-pion system 
and the recoil nucleon in the lab frame. (Blue) triangles: data for $\gamma p\rightarrow \pi^0\pi^0 p(n)$, (red)  
inverted triangles: data for $\gamma n\rightarrow \pi^0\pi^0 n(p)$, dashed (green) lines: 
MC simulations for $\pi^0\pi^0$ production, dotted (magenta) lines: MC for background reactions, 
solid (black) lines: sum of both. Dashed vertical lines: applied cuts (1.5$\sigma$).
}
\label{fig:cop}       
\end{figure*}
\begin{figure*}[Htb]
\centerline{\resizebox{1.0\textwidth}{!}{%
  \includegraphics{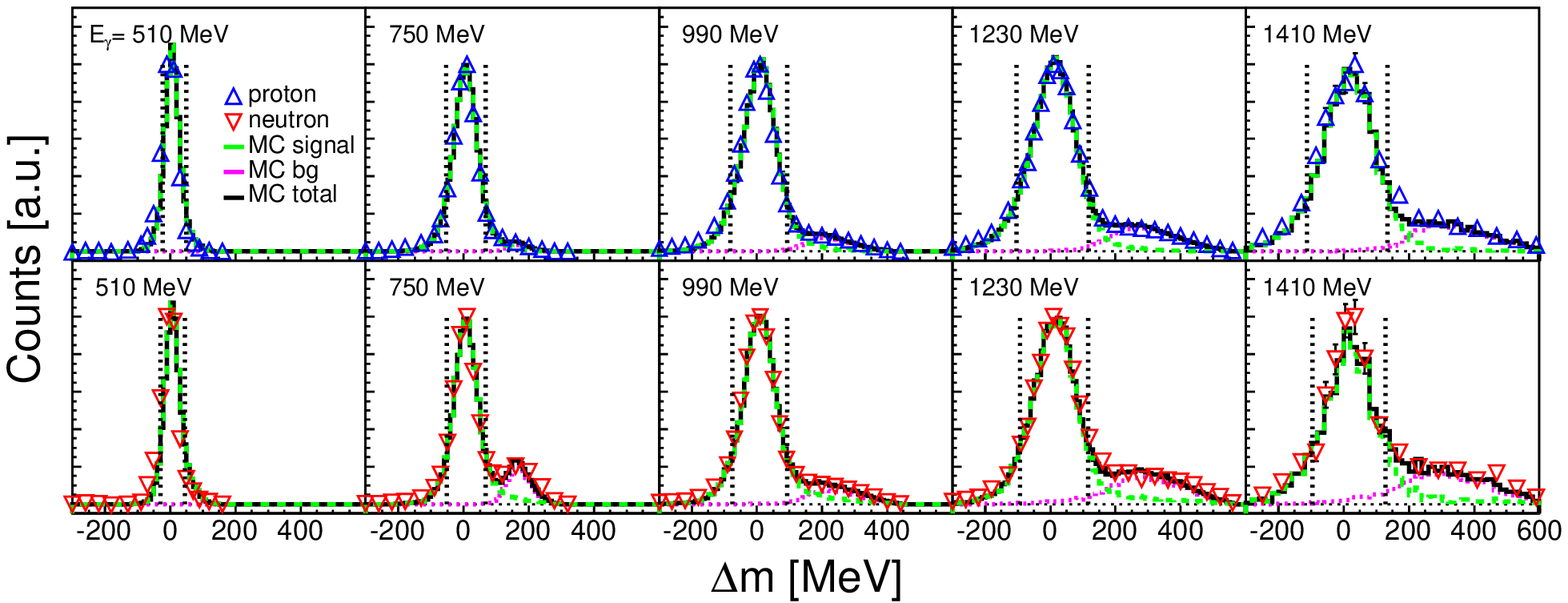}
}}
\caption{Missing-mass distributions for quasi-free $2\pi^0$ production. Notation as in 
Fig.~\ref{fig:cop}. The background peak around $\Delta m\approx$~200~MeV for the neutron data
in the 750~MeV photon-energy range is due to $\eta\rightarrow 3\pi^0\rightarrow 6\gamma$
decays where recoil neutron and one decay photon have escaped detection.
  }
\label{fig:mm}       
\end{figure*}

Since the resolution of the detector system is worse for energies than for angles, the nominal
pion invariant mass $m_{\pi^0}$ was used to correct the measured photon energies $E_i$ via 
\begin{equation}
E'_{i}\ =\ E_{i}\frac{m_{\pi^{0}}}{m_{\gamma\gamma}}\;\; {\rm with}\;\;\; i=1,2,  
\end{equation}
where $m_{\gamma \gamma}$ are the measured invariant masses. Using the corrected energies $E'_{i}$
improved the resolution for the following missing-mass analysis. 

The last step of the reaction identification was the removal of residual events from final states 
with higher pion multiplicity. For example, when a charged, low-energy pion was stopped before it 
reached the detector or was emitted too close to the beam axis, the residual events can leak into 
the data. Such events were suppressed with a co-planarity and a missing-mass analysis.  
In the center-of-momentum (cm) frame the two-pion system and the recoil nucleon are emitted 
back-to-back. Therefore, the difference between the azimuthal angle of the three-momentum vector of 
the nucleon and the sum of the momentum vectors of the two pions must be 180$^{\circ}$ (co-planar). 
Typical angular difference spectra for different incident photon energies and $\pi^0\pi^0 p$, 
$\pi^0\pi^0 n$ final states are shown in Fig.~\ref{fig:cop} and compared to the results of Monte Carlo 
simulations. Measured data and simulated line shapes are in good agreement and the background 
(mainly from $\eta\pi^0\rightarrow 4\gamma$ and $\eta\rightarrow 3\pi^0\rightarrow 6\gamma$) 
is low. It is more significant for the $\pi^0\pi^0 n$ final state due to combinatorial `self-background' 
from events where a photon was misidentified as a neutron and vice versa. Events in the range between 
the dotted vertical lines in Fig.~\ref{fig:cop} were accepted for further analysis.

Residual background below the co-planarity peaks was removed by a missing-mass analysis,
treating the recoil nucleons, although detected, as missing particles. The missing mass 
was calculated from reaction kinematics as
\begin{equation}
\label{eq:2pimiss}
\Delta m(\pi\pi) = \left|P_{\gamma}+P_{N}- P_{\pi^0_1} -P_{\pi^0_2}\right| -m_N\ ,
\end{equation}
where $m_N$ is the nucleon mass, $P_{\gamma}$ is the four-momentum of the incident photon, 
$P_N$ is the four momentum of the initial state nucleon (assumed at rest), and $P_{\pi^0_{1,2}}$ 
are the four momenta of the $\pi^0$ mesons. Typical missing mass distributions are shown in 
Fig. \ref{fig:mm} (cut on co-planarity was already applied). The line shapes are well reproduced 
by Monte Carlo simulations. Finally, events were accepted within $\pm 1.5\sigma$ of the peak position. 
Possible residual background 
(again from $\eta\pi^0\rightarrow 4\gamma$ and $\eta\rightarrow 3\pi^0\rightarrow 6\gamma$)
in this region appears to be low and its shape agrees with the MC simulations. 

The missing mass and the co-planarity peaks are broadened by the Fermi motion of the 
nucleons bound in the deuteron. This was taken into account in the event generator
for the Monte Carlo simulations. The spectra in Figs.~\ref{fig:cop} and \ref{fig:mm} are all for the 
deuterium target. The peaks are more narrow for the measurements with the hydrogen target.

After the application of the kinematic cuts, the particle identification spectra ($\Delta E-E$
for CB-PID in Fig.~\ref{fig:ede}, PSA in Fig~\ref{fig:tpsa}, and ToF-versus-energy for TAPS in 
Fig~\ref{fig:ttof}) were checked again for residual background. Only the indicated cuts for the 
PSA spectra were applied in the final analysis of the data. The $\Delta E-E$ PID-CB spectrum 
(see Fig.~\ref{fig:ede}) showed no contamination of the proton band with charged pions. 
The ToF-versus-energy spectra for TAPS (see Fig.~\ref{fig:ttof}) showed no contamination of the 
photon bands with massive particles, and the neutron candidates showed no traces of residual 
protons. In the previous analysis of the same data set for other final states, $\Delta E-E$ 
spectra for TAPS-CPV (lower resolution than for CB-PID), and cluster-multiplicity distributions 
for photons and neutrons in the CB were also analyzed (see \cite{Werthmueller_14}), all without any 
traces of significant background. 
 
Due to the high incident photon flux, several tagger channels usually responded
for each event detected in the calorimeter (typical values were distributed around a mean of 35 hits 
in the tagger focal plane per event). Most of the resulting random coincidences were removed by a cut 
on the tagger - calorimeter coincidence timing (resolution FWHM: 0.9~ns TAPS versus tagger, 1.5~ns 
CB versus tagger). The background below the coincidence peak was reduced by the missing-mass cut 
(because for randomly coincident hits the incident photon energy is not correct). The residual 
random background was subtracted in the usual way by a side-band analysis of the timing spectrum 
(see \cite{Werthmueller_14} for details).   

For the reactions off the free proton, the relevant center-of-momentum (cm) energy $W=\sqrt{s}$ is directly 
related to the incident photon energy $E_{\gamma}$ by
\begin{equation}
W = \sqrt{2E_{\gamma}m_p + m_p^2}
\end{equation} 
where $m_p$ is the mass of the proton. For the deuterium target, this relation is only approximate 
because of nuclear Fermi motion. However, the effective $W$ of the  final state can be reconstructed 
from the fully determined reaction kinematics such that the Fermi-motion effects can be eliminated. 
The reaction kinematics is determined by the incident photon energy, the rest masses of the involved 
particles (deuterons, nucleons, and pions), the final-state momenta of the two pions, and the polar 
and azimuthal angles of the participant nucleon using the four constraints from energy and momentum 
conservation. The kinetic energy of the recoil nucleon, which in general is not measured for recoil 
neutrons, is not needed (see \cite{Werthmueller_14,Krusche_11,Jaegle_11}) for details. 
This reconstruction has been well tested and worked even (with some reasonable approximations) for 
the more complicated case of a $^3$He target nucleus \cite{Witthauer_13}. 

The absolute normalization of the cross-section data follows from the target densities 
(see Table~\ref{tab:beam}), the incident photon flux, and the detection efficiency. The photon
flux was already determined for the analysis of other reaction channels
\cite{Dieterle_14,Werthmueller_14} from the same data. This was based on the count of the 
scattered electrons by the tagger focal plane scalers and periodic measurements of the tagging 
efficiency, i.e. the fraction of tagged photons that pass through the collimator. The tagging efficiency
was repeatedly measured at strongly reduced beam intensity with a detector moved into the photon 
beam line downstream of the target. For the analysis of cross sections as a function of the 
kinematically reconstructed $W$, the photon flux distribution had to be folded in with the momentum 
distribution of the bound nucleons (taken from \cite{Lacombe_81}), which was done in this analysis
as discussed in detail in \cite{Werthmueller_14}.

The most critical ingredient was the determination of the detection efficiency. It was mainly
based on Geant4 Monte Carlo simulations \cite{Geant4} complemented with direct 
measurements of the recoil nucleon detection efficiencies. In the first step the reactions
were simulated using event generators that included the momentum distributions of the bound 
nucleons and were based on different assumptions about the reaction mechanism (see next paragraph).
From these simulations, the detection efficiencies for the $\pi^0\pi^0 p$ and $\pi^0\pi^0 n$
final states were extracted as a function of incident photon energy $E_{\gamma}$ (respectively 
of reconstructed $W$) and the cm polar angle of the pion-pion system (back-to-back with the 
recoil nucleon). 
\begin{figure}[thb]
\centerline{\resizebox{0.47\textwidth}{!}{%
  \includegraphics{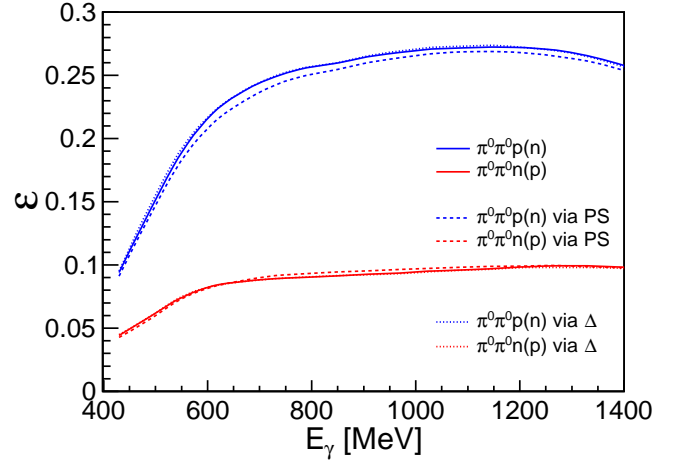}
}}
\caption{Total detection efficiency for $\pi^0\pi^0 p(n)$ (quasi-free protons from deuterium target, 
blue curves), and $\pi^0\pi^0 n(p)$ (quasi-free neutrons, red curves). The dotted curves (almost 
indistinguishable from the solid curves) correspond to the 
$\gamma N\rightarrow N^{\star},\Delta^{\star}\rightarrow \pi^0\Delta(1232)3/2^+\rightarrow \pi^0\pi^0 N$
decay chains (I), the dashed curves to phase-space (II), 
and the solid curves to the final weighted total efficiency (see text).
}
\label{fig:epsit}       
\end{figure}
Since the Geant4 code is well tested and reliable for electromagnetic showers 
but not for low-energy recoil nucleons, detection efficiencies for the latter were also 
investigated experimentally. For this, the reactions $\gamma p\rightarrow \pi^0\pi^0 p$ and 
$\gamma p\rightarrow \pi^+\pi^0 n$ from the free proton target were used. The nucleon detection 
efficiencies were simply determined from the numbers of $\pi^0\pi^0$ or $\pi^0\pi^+$ 
events with and without coincident recoil nucleons. Subsequently, these reactions were simulated 
with Geant4 and recoil nucleon detection efficiencies were extracted from the simulated data in 
the same way. The ratios of simulated and measured nucleon detection efficiencies as a function 
of the laboratory polar angle and kinetic energy of the recoil nucleons were then used to correct 
the results from the Monte Carlo simulations for the deuterium targets. Note that the relevant energy 
thresholds for the detection of recoil nucleons were not the 30 MeV (CB), respectively 35 MeV (TAPS) 
trigger thresholds because the trigger was generated by the decay photons. The total summed-up 
cluster energy for neutron-hit candidates had only to pass a 20~MeV software threshold.

\begin{figure}[thb]
\centerline{\resizebox{0.47\textwidth}{!}{%
  \includegraphics{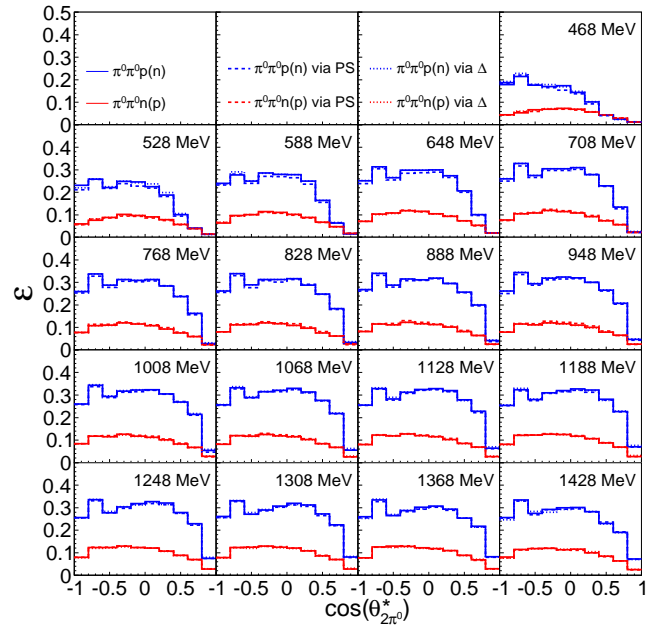}
}}
\caption{Detection efficiency for different bins of incident photon energy as function of
$\Theta^{\star}_{2\pi^0}$. Notation for curves as in Fig.~\ref{fig:epsit}.
}
\label{fig:epsid}       
\end{figure}

Since the simulations of the detection efficiency could not be done in dependence of a complete 
set of independent kinematic variables (which would require a five-dimensional space for which 
the statistical quality of the data was not sufficient), they depend in principle on the
choice of the event generator. Most critical in this aspect are the kinetic energy distributions 
of the recoil nucleons and their (correlated) angular distributions. The efficiency dependence 
on the pion distributions is rather flat due to their two-photon decays, which average for a given 
pion kinematics over many detector properties. The efficiencies were therefore simulated with 
different event generators (reflecting the dominant processes discussed in Sec.~\ref{sec:results}). 
One generator used three-body phase space (I) for the $\pi^0\pi^0 N$ final state, a second (II) 
modeled the decay chains
$\gamma N\rightarrow N^{\star},\Delta^{\star}\rightarrow \pi^0\Delta(1232)3/2^+\rightarrow \pi^0\pi^0 N$,
and the third (III) (only important for the highest incident photon energies) simulated the decay chain 
$\gamma N\rightarrow N^{\star},\Delta^{\star}\rightarrow \pi^0 N(1520)3/2^-\rightarrow \pi^0\pi^0 N$.
The results of these simulations did not show much difference. 
Figures \ref{fig:epsit} and \ref{fig:epsid} show the total and angle differential (cm polar angle 
$\Theta^{\star}_{2\pi^0}$ of the $\pi^0\pi^0$ system) detection efficiencies for event 
generators (I) and (II) and the finally used detection efficiency, which was calculated from 
the weighted average of the three event generators. The weight factors were the relative 
contributions of the three dominant reaction mechanisms determined from a combined fit of the 
simulated line-shapes to the invariant-mass distributions of the $\pi^0\pi^0$ and $\pi^0 N$ pairs 
(see Fig.~\ref{fig:tot_fractions}). 
These efficiencies already include the corrections for experimentally determined recoil nucleon 
detection efficiencies. Their angular dependence is dominated by the recoil nucleons. In particular, 
for recoil protons it decreases strongly for forward angles of the two-pion system. Such events 
correspond to recoil nucleons emitted in the cm system at backward angles. They have small 
kinetic energies in the laboratory system such that the probability is high that they are stopped 
before they reach the detector. 

The above analysis leads to cross-section data for the quasi-free production of protons and neutrons 
bound in the deuteron. An estimate for the free-neutron cross section requires the elimination of
nuclear FSI effects. In the absence of detailed model calculations for such effects, one can
only make the approximation that the FSI effects are similar for incident protons and neutrons. 
In that case, one can correct them via:
\begin{equation} 
\sigma_{\rm f}(\gamma n) = 
\sigma_{\rm qf}(\gamma n) \times\frac{\sigma_{\rm f}(\gamma p)}{\sigma_{\rm qf}(\gamma p)},
\label{eq:fsi}
\end{equation}
where $\sigma_{\rm f}$ and $\sigma_{\rm qf}$ are free and quasi-free cross sections for the initial 
states $\gamma p$ and $\gamma n$, respectively. Such an approximation was also applied for 
the previous data from the GRAAL experiment \cite{Ajaka_07}. Model results for other reactions channels 
\cite{Tarasov_11,Tarasov_15} indicate that this approximation is good except at extreme forward 
angles of the meson system (which result in small relative momenta between the two final-state nucleons).
However, the quasi-free data are folded in with Fermi motion (or when kinematic final-state reconstruction 
is used with experimental resolution). This leads to artificial structures in the 
$\sigma_{\rm f}(\gamma p)/\sigma_{\rm qf}(\gamma p)$ ratio. The two peak-like structures in
the free proton excitation function are broadened for the quasi-free data (see e.g. Fig.~\ref{fig:tot_p} 
in Sec.~\ref{sec:results}), so that the correction factors are overestimated in the peak positions 
and underestimated in the valley between them. This problem can be avoided when the free cross 
section $\sigma_{\rm f}(\gamma p)$ is also folded with Fermi motion or experimental resolution. 
This procedure was applied in the present analysis but not for the previous GRAAL data
\cite{Ajaka_07}. 

\subsection{Systematic uncertainties}

The systematic uncertainties can be split into three different types. The overall normalization 
uncertainties (referred to as type (a)) affect all data in the same way (and are 
identical to those given in \cite{Kaeser_15,Werthmueller_14}). The target surface density is 
related to the measurement of the length and exact geometric shape of the target cylinder and 
the measurement of the target pressure. Typical values and uncertainties are given in Table~\ref{tab:beam}. 
Additional effects might arise from deformations of the target windows in the cooled state. 
A conservative overall uncertainty of 4\% is estimated. The uncertainty of the photon flux 
contributes 3\%. We include here also the empty target subtraction with a conservative estimate of
2.5\% (the total correction is on the order of 5\%), while the uncertainty of the 
$\pi^0\rightarrow\gamma\gamma$ decay branching ratio is negligible \cite{PDG}. 
The overall uncertainty from the above sources is between 6\% and 9\% (quadratic or linear 
addition).

The uncertainties (referred to as type (b)), which depend on reaction types and on $W$ and cm polar angles, 
but not on the absolute detection efficiency of the recoil nucleons, originate from the analysis cuts 
(particle identification, invariant mass, co-planarity, and missing mass) and the correct reflection of 
these cuts and the detector properties in the simulations of the detection efficiency. Uncertainties 
from the analysis cuts were investigated by varying them within reasonable limits; effects arising 
from the choice of the event generator (see discussion above) by a comparison of simulations with 
different reaction mechanisms. In total, uncertainties in the 5\% - 10\% range were estimated 
(depending on incident photon energies, pion angles, and invariant masses).

The most serious source of systematic uncertainty is the recoil nucleon detection efficiency,
which directly affects the ratio of neutron and proton cross sections. Estimates based on the comparison 
of MC simulations and direct measurements of the detection efficiency (which have been used to improve 
the MC results) indicate maximum uncertainties at the 10\% level. An independent cross check of this 
uncertainty can be done by comparing the results from the quasi-free measurements in coincidence 
with recoil nucleons ($\sigma_{\rm qfp}$, $\sigma_{\rm qfn}$) to the results from the fully inclusive 
analysis ($\sigma_{\rm incl}$) where all $\pi^0\pi^0$ events with and without recoil nucleon detection 
were accepted. The latter does not depend on nucleon detection efficiencies. They must be related by
the following equation: 
\begin{equation}
\sigma_{\rm incl} =  \sigma_{\rm qfp} + \sigma_{\rm qfn} , 
\label{eq:incl}
\end{equation}
where contributions from the coherent $\gamma d\rightarrow \pi^0\pi^0 d$ reaction are negligible, 
which was estimated in \cite{Egorov_15} well below the 100 nb level and preliminary analyses 
of the present data support this result. As shown in Fig.~\ref{fig:tot_np}, the present results above 
$W$=1500~MeV have deviations that are below 3\% and also the analyses of single $\eta$ production 
\cite{Werthmueller_14}, of single $\pi^0$ production \cite{Dieterle_14}, and of $\pi\eta$-pairs 
\cite{Kaeser_15} from the same data set are in good agreement with Eq.~\ref{eq:incl}, typically 
better than within 5\%.

A further test comes from a comparison of the results for the free-proton data analyzed with and 
without coincident detection of the recoil protons (see Fig.~\ref{fig:tot_p} in Sec.~\ref{sec:results}), 
which also agree within 5\% (for most of the energy range better than $\pm$3\%, with the average 
agreement over the whole energy range better than 0.5\%). Therefore, the 10\% estimate is very 
conservative, apart from the threshold region.   

The above uncertainties apply to the directly measured quasi-free cross sections. However, more 
relevant is the systematic uncertainty of the best approximation for the cross section
for a `free'-neutron target. Due to Eq.~(\ref{eq:fsi}), absolute normalization uncertainties
(type (a)) for the quasi-free measurements cancel. Also, uncertainties due to analysis cuts (type (b)) 
cancel almost completely. Therefore, remaining major sources for systematic uncertainty are the detection 
efficiencies for the recoil nucleons (type (c)), the effects from Fermi motion and nuclear FSI effects, 
and the absolute normalization of the free-proton data. The influence of Fermi motion was investigated 
by the comparison of results from the analyses constructing $W$ from the initial state (`IS', influenced 
by Fermi motion) and from the final state (`FS', independent of Fermi motion; see Sec.~\ref{sec:results}). 
The nuclear FSI effects were studied by the comparison of free and quasi-free proton data 
(see Sec.~\ref{sec:results}) and are substantial. However, what matters is the difference of such 
effects for recoil protons and neutrons (see discussion above), which is expected to be small except 
for certain extreme kinematics. Therefore, only the conservative 10\% estimate from the type (c) effects 
dominates. The systematic type (a) and type (b) uncertainties of the free-proton data are comparable to 
the measurements with the deuterium data. Additional uncertainties from proton detection were not 
observed for the free proton data.

\section{Results}
\label{sec:results}
\begin{figure}[thb]
\centerline{\resizebox{0.5\textwidth}{!}{%
  \includegraphics{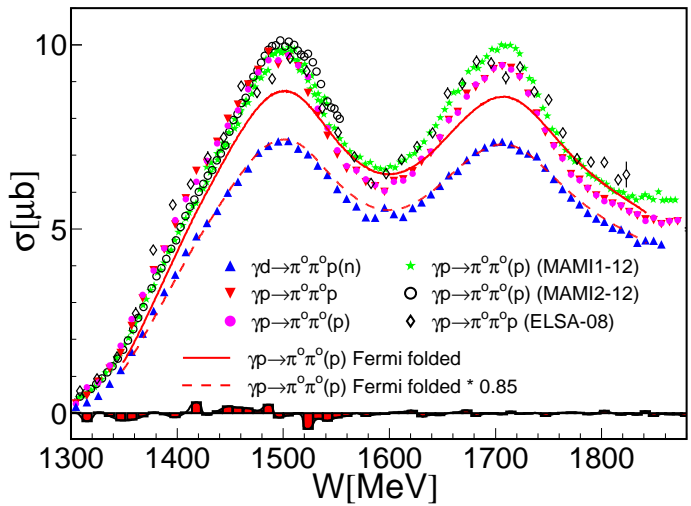}
}}
\caption{Total cross section for the $\pi^0\pi^0 p$ final state as a function
of $W(E_{\gamma})$ (`IS' analysis). Present results for a free-proton target from analyses with 
(red downward triangles) and without (magenta dots) coincident detection of the 
recoil protons. Previous results (ELSA-08: \cite{Thoma_08}, MAMI1-12: \cite{Kashevarov_12}, 
MAMI2-12: \cite{Zehr_12}). Present results for quasi-free protons (blue upward triangles). 
Histogram at the bottom: difference between the present analyses with and without coincident
recoil protons for free protons. Solid (dashed) curves: free-proton cross section folded 
with Fermi motion (scaled by factor 0.85).
Legend: nucleons without brackets are detected in coincidence, nucleons in brackets are 
not required.
}
\label{fig:tot_p}       
\end{figure}

Total cross-section data for the $\gamma p\rightarrow \pi^0\pi^0 p$ reaction are summarized in 
Fig.~\ref{fig:tot_p}. The measurement with the liquid hydrogen target (free protons) was analyzed in two 
different ways: with and without requiring coincident detection of the recoil protons. The two analyses
are in good agreement (see bar histogram at the bottom of Fig.~\ref{fig:tot_p} for the difference). 
In Fig.~\ref{fig:tot_p}, the free-proton cross-section data are compared to previous results obtained 
at ELSA \cite{Thoma_08} (with coincident proton detection) and MAMI \cite{Kashevarov_12,Zehr_12} 
(without coincident proton detection). Overall, the agreement is satisfactory, with the largest discrepancies 
observed in the high-energy tail of the third resonance region (on the order of 10\% to the MAMI data).
Both analyses of the present free-proton data agree with previous measurements 
within systematic uncertainties (statistical uncertainties are almost negligible except in the 
vicinity of the threshold).

\begin{figure}[htb]
\centerline{\resizebox{0.50\textwidth}{!}{%
  \includegraphics{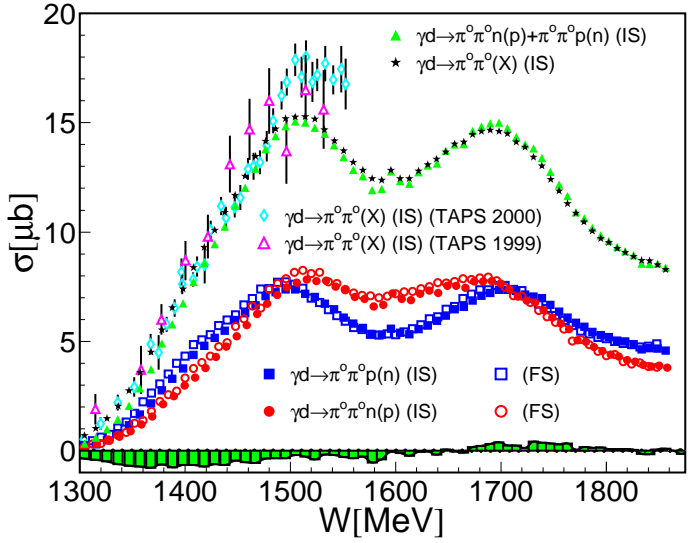}
}}
\caption{Total cross sections for $\gamma d\rightarrow \pi^0\pi^0 p(n)$ (blue squares),  
$\gamma d\rightarrow \pi^0\pi^0 n(p)$ (red circles) and the sum of both (green triangles)
as a function of $W$. Open symbols: `IS' analysis, filled symbols: `FS' analysis (see text). 
(Black) stars: fully inclusive analysis. (Green) histogram: difference between `IS' and `FS'
analysis. Previous inclusive results: (cyan) diamonds: \cite{Kleber_00}, (magenta) open triangles: 
\cite{Krusche_99}.
}
\label{fig:tot_np}       
\end{figure}

\begin{figure}[htb]
\centerline{\resizebox{0.5\textwidth}{!}{%
  \includegraphics{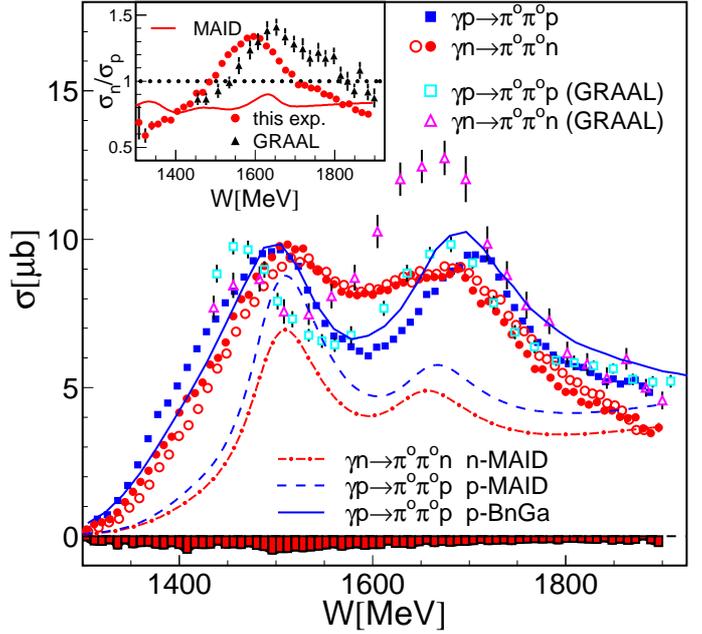}
}}
\caption{Main plot: total cross section for $\gamma p\rightarrow \pi^0\pi^0 p$ (free-proton data) from
present experiment (blue filled squares) and GRAAL (open cyan squares) \cite{Assafiri_03}.
Present results for `free' neutron target: red filled (open) circles from FS (IS) analysis.
Histogram at bottom: systematic uncertainty of `free neutron' data excluding overall normalization
uncertainty (of free proton data).
Previous `free neutron' data from GRAAL \cite{Ajaka_07} (open magenta triangles). 
Dashed (blue) and dash-dotted (red) curves: results of the MAID model \cite{Fix_05} for proton and neutron 
targets, respectively. 
Solid (blue) curve: BnGa model \cite{Anisovich_12} for proton. Insert: ratios of neutron and 
proton quasi-free cross sections for present (red circles) and GRAAL (black triangles) data.
}
\label{fig:tot_final}       
\end{figure}

The free-proton data are also compared in Fig.~\ref{fig:tot_p} to the present results for the 
quasi-free reaction $\gamma d\rightarrow \pi^0\pi^0 p(n)$ (detected participant proton and 
undetected spectator neutron). The data shown in Fig.~\ref{fig:tot_p} are based on $W$ and extracted 
from the initial state (`IS' analysis), like the free-proton data and thus subject to Fermi 
smearing. The size of this effect can be estimated by folding the free-proton data with the 
momentum distribution of the bound nucleons calculated from the deuteron wave function \cite{Lacombe_81}. 
The result is the solid curve shown in Fig.~\ref{fig:tot_p}. It overestimates the measured quasi-free
data by approximately 15\% independent of incident photon energy (the dashed curve is
down scaled by a factor of 0.85 and agrees well with the data). Although the estimated 
systematic uncertainty of the quasi-free cross sections is of the same magnitude, it is  
unlikely that this discrepancy is due to the absolute calibration of the data. 
The results for other reaction channels extracted from the same data set show different types of
behavior. For single $\pi^0$ production \cite{Dieterle_14}, free and quasi-free proton data 
differ up to 35\% and the effect is energy dependent. For $\eta\pi^0$ pairs \cite{Kaeser_15},
the discrepancy is around 30\%, for $\eta\pi^+$ pairs \cite{Kaeser_15} around 10\%, and for 
single $\eta$ production \cite{Werthmueller_13,Werthmueller_14}, free and quasi-free results 
are in almost perfect agreement. The most probable explanation is that there is a strong nuclear FSI 
that depends on the reaction channel. The general pattern of the effects is more or less as 
expected. They are known to be larger for neutral pions than for charged pions \cite{Krusche_03}
(the nucleon - nucleon FSI is different for proton - neutron and neutron - neutron pairs).  
Furthermore, many experiments \cite{Krusche_03} have shown that they are almost negligible for single
$\eta$ production (mostly due to the dominance of the excitation of $s$-channel resonances, which 
enforce a nucleon spin-flip). A detailed understanding of these effects is still lacking, although 
there are some recent model results for the $\gamma d\rightarrow \pi^- p(p)$ \cite{Tarasov_11} and 
$\gamma d\rightarrow \pi^0 p(n)$ \cite{Tarasov_15} channels. 

Figure~\ref{fig:tot_np} summarizes the results of the `IS' and `FS' analyses of the quasi-free 
proton and neutron data and compares the sum of the two exclusive cross sections to the 
total inclusive cross section obtained from an analysis that ignores the recoil nucleons. 
For the inclusive $\gamma d\rightarrow\pi^0\pi^0 (X)$ reaction, $X$ can be a proton, or a neutron, 
or may be absent and a missing-mass analysis can be used to eliminate events with additional mesons.
The `IS' and `FS' results are similar because the elimination of the Fermi smearing effects 
is partly counteracted by experimental resolution effects in the `FS' analysis. 

The good agreement of the sum of the exclusive cross sections with the inclusive 
result demonstrates that no major sources of systematic uncertainty are related to the
detection of the recoil nucleons. Results for the inclusive cross section up to the second
resonance peak have been reported from two previous low-statistics measurements
\cite{Krusche_99,Kleber_00}. Results from the older experiment \cite{Krusche_99} agree with 
the present results within its relatively large uncertainties. The data from the second measurement 
\cite{Kleber_00} agree up to $W\approx$~1480~MeV, but they are systematically higher (roughly 13\%)
above the $\eta$ threshold. In this energy range both previous experiments had a large background 
from the $\eta\rightarrow 3\pi^0$ decay, which is absent in the present data due to the almost
$4\pi$ coverage of the detector. Altogether, no serious discrepancies were observed.

The estimate for the free-neutron cross section derived from Eq.~\ref{eq:fsi} is shown in 
Fig.~\ref{fig:tot_final}. Results are given for the `IS' and `FS' analysis of the data, 
which are quite similar. This is partly so because the Fermi smearing effects in the `IS' analysis
and the resolution effects for the `FS' analysis are of similar size and partly because, due to the
absence of pronounced peak structures, both effects are small for the neutron excitation function.
The results are compared to the free-proton cross section from the present analysis. The insert in 
Fig.~\ref{fig:tot_final} shows the ratio of neutron and proton cross sections, which was directly 
computed from the measured quasi-free cross sections. This result is free from all normalization 
uncertainties. Also shown are the data from the previous measurements of the GRAAL experiment 
\cite{Assafiri_03,Ajaka_07}, which so far provided the only exclusive results for the neutron cross 
section. 

\begin{figure*}[thb]
\centerline{\resizebox{1.0\textwidth}{!}{%
  \includegraphics{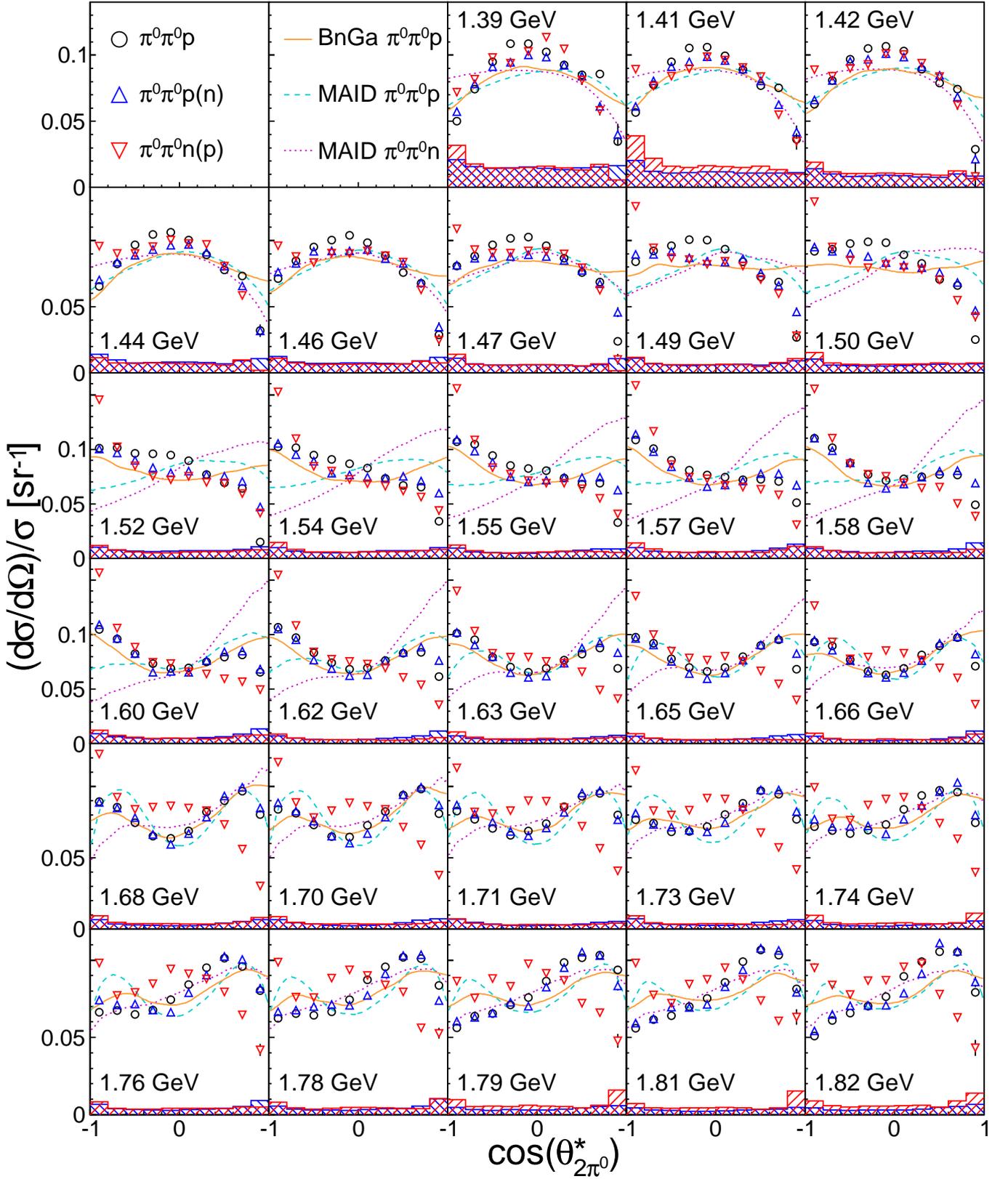}
}}
\caption{Angular distributions for free and quasi-free protons and quasi-free neutrons 
as a function of $\cos(\theta^*_{2\pi^0})$ normalized to the total cross section for different 
center-of-mass energy bins of $\pm$4~MeV (for lack of space every second bin is omitted). 
Notation given in the figure. Model predictions: full (orange)
curve: proton calculations from BnGa (\cite{Anisovich_12}), dashed (cyan) curve proton and 
dotted (magenta) curve neutron calculations from MAID (\cite{Fix_05}). Blue (red) histograms at bottom
indicate systematic uncertainties for the proton (neutron) data.
}
\label{fig:angle}       
\end{figure*} 
\clearpage

\begin{figure*}[thb]
\centerline{\resizebox{1.0\textwidth}{!}{%
  \includegraphics{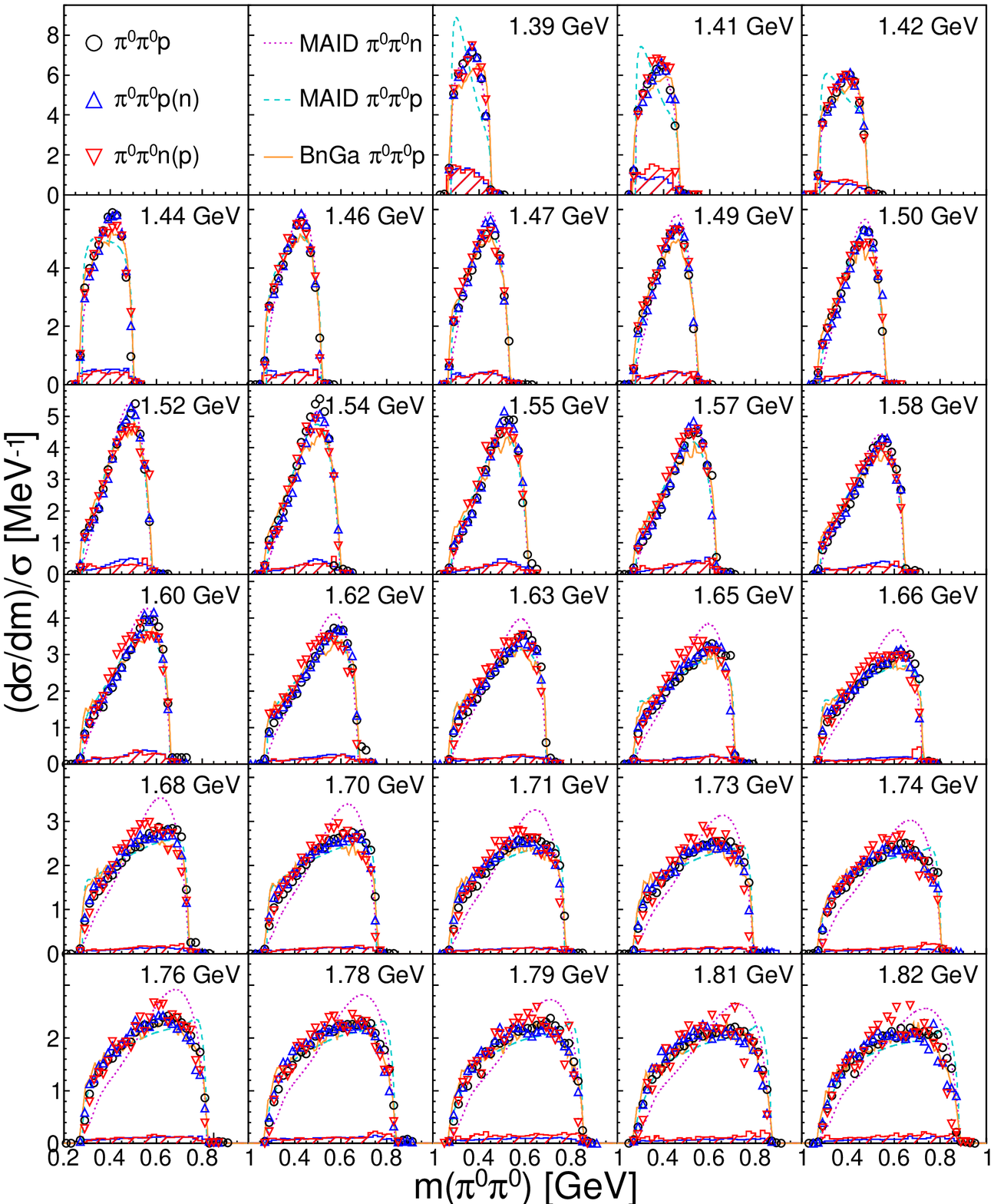}
}}
\caption{Invariant-mass distributions of $\pi^0\pi^0$ pairs for different cm energy bins. Notation as 
in Fig. \ref{fig:angle}.
}
\label{fig:m_pipi}       
\end{figure*}
\clearpage

\begin{figure*}[htb]
\centerline{\resizebox{1.0\textwidth}{!}{%
  \includegraphics{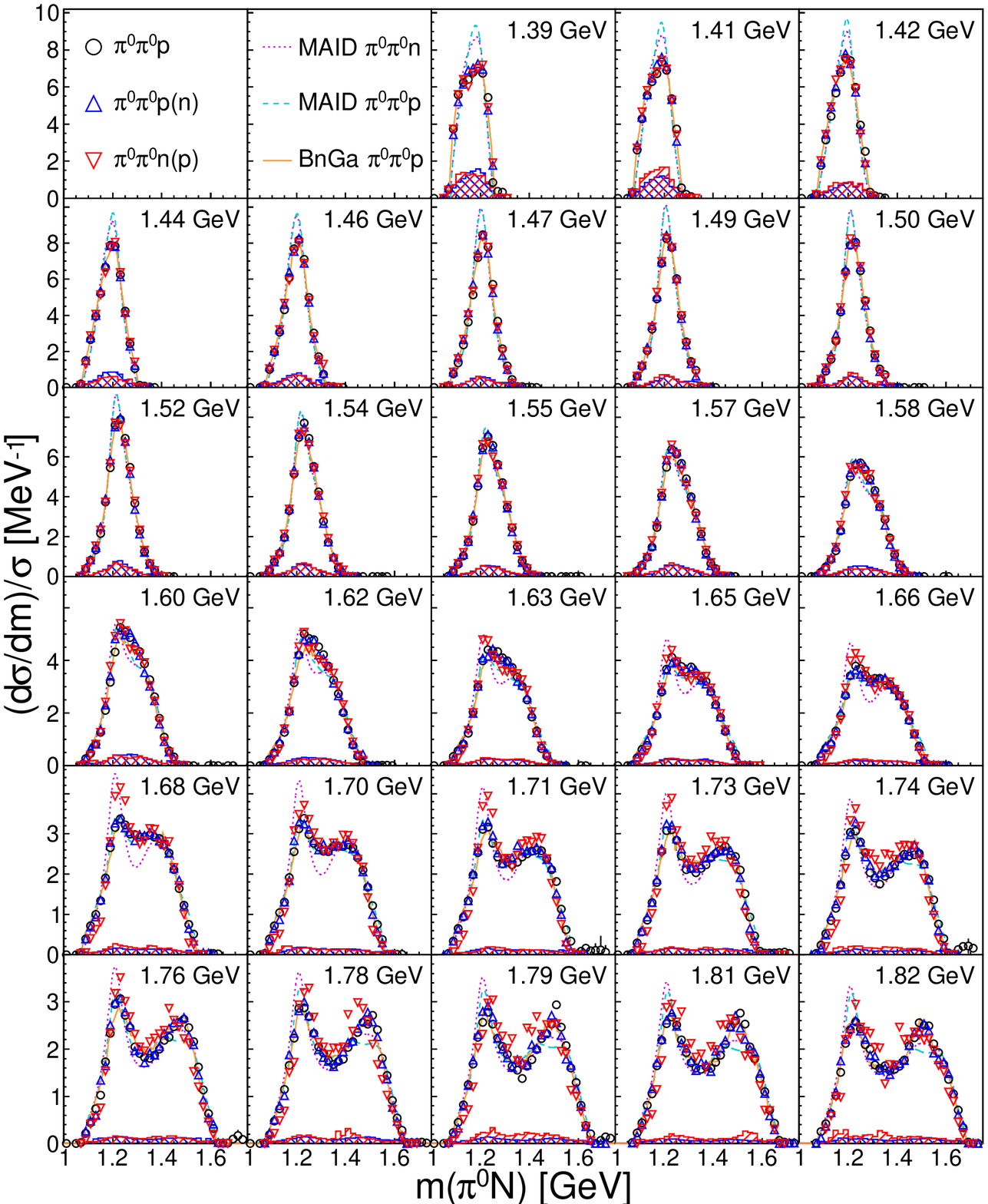}
}}
\caption{Invariant-mass distributions of $\pi^0 N$ pairs for different cm energy bins. Notation as in Fig. 
\ref{fig:angle}.
}
\label{fig:m_Npi0}       
\end{figure*}
\clearpage

\begin{figure*}[thb]
\centerline{\resizebox{0.9\textwidth}{!}{%
  \includegraphics{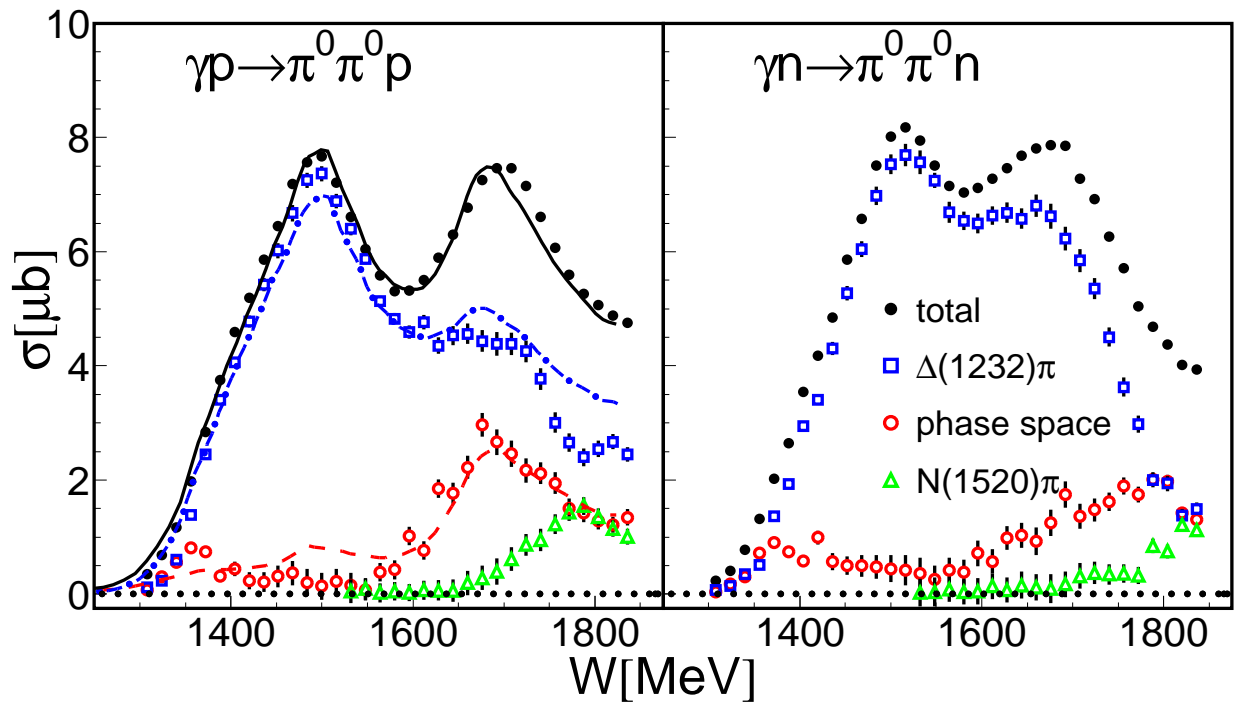}
}}
\caption{Fitted contributions of the $\gamma N\rightarrow \pi^0\Delta(1232)3/2^+\rightarrow \pi^0\pi^0 N$
reaction chain (open blue squares), phase-space decays (open red circles), and 
$\gamma N\rightarrow \pi^0 N(1520)3/2^-\rightarrow \pi^0\pi^0 N$ (open green triangles) to the
total cross section (filled black circles). Left hand side for the proton target, right hand side 
for the neutron target. Only for proton target: solid (black), dash-dotted (blue), and dashed (red)
curves represent total cross section, sequential decay over $\Delta(1232)3/2^+\pi$ intermediate state,
and phase-space component, respectively from a PWA of previous ELSA proton data 
\cite{Thoma_08}. All curves renormalized to the total cross section to account for FSI. 
}
\label{fig:tot_fractions}       
\end{figure*}

They were also obtained with Eq.~(\ref{eq:fsi}) from free and quasi-free GRAAL data with an 
`IS' analysis, but without considering the Fermi smearing in the quasi-free proton data. 
The GRAAL experiment reported a large enhancement of the neutron cross section with respect to the 
proton data in the third resonance region, but the present data do not support this. 
There are, however, significant differences between proton and neutron data in particular 
in the tails and between the two resonance peaks, which points to different reaction contributions
that are discussed below. 

Also shown are the results from two reaction models, the BnGa partial-wave analysis (PWA)
\cite{Anisovich_12} and the Two-Pion MAID model \cite{Fix_05}. It is obvious that the MAID model
is missing some important contributions, in particular in the threshold region and in the third
resonance peak. The BnGa model has been fitted to the ELSA free-proton data \cite{Thoma_08} 
and thus agrees more or less with the proton cross section. Results for the neutron target from 
this model are not yet available.

Differential cross sections were extracted only from the `FS' analysis because
variables such as cm angles or invariant masses of particle pairs are smeared too much
by Fermi motion in the `IS' analysis. In order to facilitate a comparison of the shapes
of the distributions from different reactions and from experiment to theory, all
results were normalized to their total cross sections.

Angular distributions as a function of the cm angle $\Theta^{\star}_{2\pi^0}$, which 
corresponds to the combined four momentum of the two-pion system, are summarized 
in Fig.~\ref{fig:angle}. The shapes of the free and 
quasi-free proton data are overall in good agreement; i.e., the suppression of the 
quasi-free cross sections due to FSI is almost independent of angle, which is somewhat 
contrary to FSI modeling. However, one should note that these data do not cover the extreme
forward angles, which is where the models predict the largest effects. The angular 
dependence of the quasi-free neutron data is similar to the proton data up to the maximum 
of the second resonance peak at $W\approx$~1540~MeV, but in the third resonance peak and 
above, they are quite different (see discussion below). The proton data are as expected 
and in reasonable agreement with the results of the BnGa model \cite{Anisovich_12} 
(which has been fitted to similar data from the ELSA facility). Agreement with the 
MAID model predictions \cite{Fix_05} is worse and this model also does  
not reproduce the absolute magnitude of the cross section. The only available prediction 
for the reaction off the neutron comes from the MAID model and is obviously in poor agreement 
with the experimental data, both in absolute scale and also in the shape of the 
angular distributions.    

Figures~\ref{fig:m_pipi} and \ref{fig:m_Npi0} show the invariant-mass distributions of
the pion-pion and pion-nucleon pairs. For the latter, the pions were randomized; i.e.,
for each event it was randomly chosen which pion was used. The result is the same as when 
the invariant masses from both pion-nucleon combinations were used in the spectra and 
the differential cross sections were renormalized by a factor of 1/2, but it avoids 
correlations in the statistical uncertainty. Again, the results from free and quasi-free proton 
data are in good agreement. Also for this differential cross section, proton and neutron data 
agree in shape up to the second resonance bump around $W\approx$~1550~MeV. At higher energies,
above 1600~MeV, the pion-neutron invariant-mass distributions show a more pronounced peak for the 
$\Delta$ resonance than for the pion-proton invariant masses. This suggests a larger
contribution from sequential decays via the $\Delta(1232)3/2^+$ intermediate state in the third
resonance peak of the neutron, which is clearly reflected in the fit results shown in 
Fig.~\ref{fig:tot_fractions}. 

For a more quantitative analysis the invariant-mass distributions were fitted with
the simulated line shapes for phase-space distributions, the 
$\gamma N\rightarrow \pi^0\Delta(1232)3/2^+\rightarrow \pi^0\pi^0 N$, and the 
$\gamma N\rightarrow \pi^0 N(1520)3/2^-\rightarrow \pi^0\pi^0 N$ sequential decays (the latter 
parametrized only a decay via an intermediate state around 1520~MeV and the quantum numbers were not
considered). The fits were done simultaneously in the pion-nucleon and pion-pion invariant-mass
distributions. This simple analysis ignores the information from the angular distributions 
about the quantum numbers of involved resonances and also the beam-helicity asymmetries published in 
\cite{Oberle_13} and it does not account for interferences between the different contributions. 
A more detailed analysis in the framework of coupled-channel PWAs is desirable, but not yet available.
Nevertheless, the results of the present analysis highlight already important aspects for
the comparison of this reaction off protons and neutrons. The results are summarized
in Fig.~\ref{fig:tot_fractions}. For the proton target, they can be compared to a PWA 
of free-proton data \cite{Thoma_08}. The results for the two dominant reaction components,
decays via an intermediate $\pi^0 \Delta(1232)3/2^+$ and phase-space contributions, agree quite well 
with the present analysis. 

Throughout the second resonance bump, proton and neutron cross sections are dominated 
by the sequential decay via an intermediate $\pi^0 \Delta(1232)3/2^+$ state (analyses of 
free proton data \cite{Sarantsev_08,Thoma_08,Wolf_00} have shown that this is mainly the
$N(1520)3/2^-\rightarrow \pi^0 \Delta(1232)3/2^+$ decay). The situation in the third resonance peak
is much different. For the neutron it is still dominated by a reaction chain via an intermediate
$\pi^0\Delta(1232)3/2^+$ state supplemented by smoothly rising contributions of phase-space decays
and sequential decays via an intermediate state in the $W$=1520~MeV region. For the proton target,
as in previous experiments, the $\pi^0 \Delta(1232)3/2^+$ intermediate state is less dominant
and a strong resonance structure is observed for the phase-space contribution. Somewhat higher
in energy, a peaking structure for transitions via an intermediate state in the second 
resonance region is also visible. This means that the peaks in the third resonance region are of 
completely different nature. The more pronounced peak for the proton is due to phase-space 
contributions, while the $\pi^0 \Delta(1232)3/2^+$ intermediate state is responsible for the more 
shallow neutron peak. 

For an interpretation of this pattern, one should first consider what physical processes are hidden
behind the phase-space contribution. The red dashed curve in Fig.~\ref{fig:tot_fractions} from
Ref.~\cite{Thoma_08} labeled `phase-space' actually corresponds in the PWA to the
$N(\pi^0\pi^0)_S$ final state, i.e. to decays with the two pions in a relative $s$-wave. This is sometimes
called the $N\sigma$ final state, where the broad scalar-isoscalar $\sigma$-meson is used
as an effective parametrization of this partial wave. In the MAID model \cite{Fix_05}, the dominant 
resonance contribution to the $\pi^0\pi^0 p$ final state in the third resonance bump comes from the 
$N(1680)5/2^+$ resonance, while for the $\pi^0\pi^0 n$ final state the $N(1675)5/2^-$ resonance dominates. 
Although the MAID model does not reproduce the data in this energy range, such a pattern would be also 
expected from the basic properties of these states listed in the RPP \cite{PDG}. 
The $N(1680)5/2^+$ state has a much larger electromagnetic coupling (all in units of $10^{-3}$GeV$^{-1/2}$) 
to the proton 
($A_{3/2}^p = 133\pm$12, $A_{1/2}^p = 15\pm$6) than to the neutron 
($A_{3/2}^n = 33\pm$9, $A_{1/2}^n = 29\pm$10). On the other hand, the electromagnetic excitation
of the $N(1675)5/2^-$ is Moorhouse suppressed \cite{Moorhouse_66} for the proton
($A_{3/2}^p = 20\pm$5, $A_{1/2}^p = 19\pm$8) and much larger for the neutron
($A_{3/2}^n = 60\pm$5, $A_{1/2}^n = 85\pm$10). The $5/2^-$ ($5/2^+$) states are listed with 
branching ratios to $\pi\Delta$ of 50 - 60\% (11$\pm$5\%). This can explain the large
$\pi\Delta$ component in the third resonance bump of the neutron. The $N5/2^{\pm}\rightarrow N\sigma$
decay is possible with $l_{N\sigma}$=2 for the $5/2^+$ state, but requires a minimum $l_{N\sigma}$=3 
in case of the $5/2^-$ state ($l_{N\sigma}$ relative angular momentum of the $N\sigma$ system). 
Therefore, a larger contribution from the $N\sigma$ decay for the $N(1680)5/2^+$ state (dominating the
reaction for the proton target) is plausible. These simple considerations (which do not include the
available information from angular distributions and polarization observables \cite{Oberle_13}) will 
have to be investigated in more detail with a combined PWA of both reactions.
  
\section{Summary and conclusion}
Precise data for the photoproduction of neutral pion pairs off nucleons have been measured.
The main findings are the following:

The comparison of free and quasi-free data for the proton shows that nuclear FSI effects are significant.
The quasi-free cross section for protons bound in the deuteron is reduced by approximately 15\% with 
respect to the free-proton cross section. This reduction is almost independent of final state 
invariant mass from threshold up to $W\approx$1.85~MeV and affects only the absolute scale of the cross
section; the shape of angular and invariant mass distributions is practically undisturbed. 
Compared to other final states, this effect is of medium size (ranging from almost no effect for 
$\eta$ production \cite{Werthmueller_13,Werthmueller_14}, over 10\% effects for $\eta\pi^{\pm}$ pairs 
\cite{Kaeser_15}, to $\approx 30\%$ for $\eta\pi^0$ pairs \cite{Kaeser_15}, and up to 35\% effects for 
single $\pi^0$ production \cite{Dieterle_14}). These effects are still awaiting a more detailed treatment 
in the framework of reaction models.

The total cross section for the neutron target shows a similar double-hump structure as the reaction 
on the proton (although the valley between the two bumps is less pronounced), but the analysis of the 
invariant-mass spectra of the pion-pion and in particular pion-nucleon pairs shows that the origin 
is different. For both nucleons, the first bump is dominated by a reaction chain involving an intermediate
$\pi^0\Delta(1232)3/2^+$ state. Analyses of the $\gamma p\rightarrow \pi^0\pi^0 p$ reaction \cite{Thoma_08}
have assigned this to the decay of the $N(1520)3/2^-$ state to the $\Delta$ resonance. 
However, while the second bump around $W\approx$~1700~MeV is mainly due to the $N\sigma$ channel for the proton, 
for the case of the neutron, this structure is dominated by a sequential decay via the $\pi^0 \Delta(1232)3/2^+$ 
intermediate state and the $N\sigma$ final state only contributes a smoothly rising component. This behavior 
can be interpreted such that the second bump for the proton is dominated by the decay of the $N(1680)5/2^+$
state, while the largest contribution for the neutron comes from the decay of the $N(1675)5/2^-$ 
resonance. At the highest values of $W$ covered by this experiment (see Fig.~\ref{fig:tot_fractions}, 
left hand side), the proton target shows also a more significant contribution from reactions involving an 
intermediate state in the $W$=1500~MeV range, most probably of the $N(1520)3/2^-$ resonance. In this context, 
it is surprising that the beam-helicity asymmetries reported in \cite{Oberle_13} are so similar for proton 
and neutron targets in the second bump and above. This seems to indicate that the dominant contributions 
to this asymmetry are either not related to the dominant resonance contributions or that the $5/2^+$ and 
$5/2^-$ states produce almost identical asymmetries. At present, the existing reaction models do not 
give much guidance. The Bonn-Gatchina analysis \cite{Anisovich_12} is only available for the proton 
target and the Two-Pion MAID model \cite{Fix_05} is in so much disagreement with the experimental data
in the second bump region that one cannot draw any conclusions. A combined PWA
of proton and neutron data is highly desirable. Further data for the polarization observables 
$E$ (circularly polarized beam, longitudinally polarized target), $T$ (transversally polarized target), 
and $F$ (circularly polarized beam and transversally polarized target) for proton and neutron target
are already under analysis and will put much tighter constraints on the reaction mechanisms. 
 
{\bf Acknowledgments}
We wish to acknowledge the outstanding support of the accelerator group 
and operators of MAMI. 
This work was supported by Schweizerischer Nationalfonds
(200020-156983,132799,121781,117601,113511), Deutsche
For\-schungs\-ge\-mein\-schaft (SFB 443), the INFN-Italy, 
the European Community-Research Infrastructure Activity under 
FP7 programme (Hadron Physics2, grant agreement No. 227431), 
the UK Science and Technology Facilities Council (ST/J000175/1
ST/G008604/1), the Natural Sciences and Engineering Research Council
(NSERC) in Canada. This material is based upon work also
supported by the U.S. Department of Energy, Office of Science,
Office of Nuclear Physics Research Division, under Award
Numbers DE-FG02-99-ER41110, DE-FG02-88ER40415, and
DE-FG02-01-ER41194 and by the National Science Foundation,
under Grant No.s PHY-1039130 and IIA-1358175.	
We thank the undergraduate students of Mount Allison University 
and The George Washington University for their assistance.

\end{document}